\documentclass[a4paper]{cas-sc}

\usepackage[numbers,compress]{natbib}
\usepackage{lineno,hyperref,amsmath,amssymb,bm,tikz,multirow}
\usepackage{graphicx}
\usepackage{float} 
\usepackage{subfigure}
\usepackage{caption}
\usepackage{amsmath}
\usepackage{array}
\usepackage{multirow}

\usepackage{array}
\usepackage{booktabs}
\usepackage{multirow}
\usepackage{adjustbox}
\usepackage[column=0]{cellspace} 
\usepackage{tabularray}

\def\tsc#1{\csdef{#1}{\textsc{\lowercase{#1}}\xspace}}
\tsc{WGM}
\tsc{QE}
\tsc{EP}
\tsc{PMS}
\tsc{BEC}
\tsc{DE}

\makeatletter
\renewcommand{\fnum@figure}{Fig. \thefigure.\@gobble}
\makeatother
\usepackage{graphicx} 
\usepackage{float} 
 \usepackage{amsmath}
 \usepackage{amssymb}
 \usepackage{amsmath}

\usepackage{booktabs}      
\usepackage{makecell}      
\usepackage{multirow}
\usepackage{graphicx}
\usepackage{adjustbox}     
\usepackage{array}         

\begin{document}

\let\WriteBookmarks\relax
\def\floatpagepagefraction{1}
\def\textpagefraction{.001}
\shorttitle{Gaussian Ensemble Topology (GET) Method for Topology Optimization }
\shortauthors{Xinyu Ma et~al.}

\title [mode = title]{Gaussian Ensemble Topology (GET): A New Explicit and Inherently Smooth Framework for Manufacture-Ready Topology Optimization} 

\author[1]{Xinyu Ma}

\credit{Conceptualization, Methodology, Investigation, Formal analysis, Validation, Programming, Writing-original draft}

\author[3]{Chengxin Wang}
\credit{Investigation, Validation, Writing-original draft}

\author[1]{Meng Wang}
\credit{Methodology, Formal analysis, Writing-original draft}

\author[4]{Xu Guo}
\credit{Supervision, Writing-review \& editing}

\author[2]{Liu Yang}[orcid=0000-0002-7476-9168]
\cormark[1]
\ead{yangliu@nus.edu.sg}
\credit{Conceptualization, Methodology, Funding acquisition, Supervision, Writing-original draft, Writing-review \& editing}

\author[1]{Huajian Gao}[orcid=0000-0002-8656-846X]
\cormark[1]
\ead{gao.huajian@tsinghua.edu.cn}
\cortext[cor1]{Corresponding author}
\credit{Conceptualization, Methodology, Funding acquisition, Supervision, Writing-review \& editing}

\address[1]{Mechano-X Institute, Applied Mechanics Laboratory, Department of Engineering Mechanics, Tsinghua University, Beijing 100084, China}

\address[2]{Department of Mathematics, National University of Singapore, Singapore 119076, Singapore}

\address[3]{State Key Laboratory of Turbulence and Complex Systems, School of Mechanics and Engineering Science, Peking University, Beijing 100871, China}

\address[4]{State Key Laboratory of Structural Analysis Optimization and CAE Software for Industrial Equipment, Department of Engineering Mechanics, Dalian University of Technology, Dalian 116024, China}

\begin{abstract}
We introduce the Gaussian Ensemble Topology (GET) method, a new explicit and manufacture-ready framework for topology optimization in which design geometries are represented as superpositions of anisotropic Gaussian functions. By combining explicit Gaussian descriptions with a level-set-like Heaviside projection, GET inherently generates smooth, curvature-continuous designs without requiring post-processing steps such as mesh or corner smoothing and feature extraction. The method is validated on standard compliance-minimization and compliant mechanism benchmarks in two and three dimensions. The optimized designs achieve objective values comparable to those obtained with classical Moving Morphable Component (MMC) approaches, but with geometrically consistent, refined boundaries. Numerical examples demonstrate additional advantages of the GET framework, including mesh independence inherent to explicit parameterizations, strong geometric expressiveness, and effective control over smoothness, discreteness, and structural complexity through parameter tuning. As a robust and manufacture-ready approach to explicit topology optimization, GET opens avenues for tackling advanced and complex design problems.
\end{abstract}


\begin{highlights}
\item A novel explicit framework, Gaussian Ensemble Topology (GET), represents geometry as a superposition of anisotropic Gaussian functions, producing inherently smooth and curvature-continuous designs without post-processing or feature extraction.
\item Demonstrated on classical compliance-minimization and compliant mechanism benchmarks in two and three dimensions, GET achieves performance comparable to the MMC method while yielding smoother and more refined structural boundaries.
\item Parametric studies show how the number of Gaussian fields, the regularization parameter $\epsilon$, and the cut-off threshold $T$ govern structural complexity, discreteness, and smoothness. 
\end{highlights}

\maketitle

\section{Introduction}

The growing demand for lightweight, high-performance structures in aerospace \citep{zhu2016topology,munk2019benefits}, automotive \citep{cavazzuti2011high,vido2024computer}, and civil engineering \citep{zegard2020advancing,stoiber2021topology} has accelerated advances in structural optimization. In parallel, modern manufacturing methods—including additive manufacturing \citep{jihong2021review,langelaar2016topology} and casting \citep{lei2018investment,almonti2025lightweight}—enable geometries that were previously impractical. Within this context, topology optimization (TO) seeks the optimal material layout in a prescribed domain to minimize a chosen objective under constraints. Since the seminal work of Bendsøe and Kikuchi \citep{bendsoe1988generating}, TO has progressed substantially across formulations and numerics, including SIMP \citep{bendsoe1989optimal,zhou1991coc,bendsoe2013topology}, level-set \citep{osher1988fronts,osher2004level, wang2003level,allaire2004structural}, ESO/BESO \citep{xie1993simple,yang1999bidirectional}, and the explicit MMC family \citep{guo2014doing,zhang2016new,du2022efficient}. Despite their success, prevalent pixel/element-based (implicit) approaches often produce jagged boundaries and gray regions requiring geometric reconstruction, while explicit MMC-type methods, though CAD-friendly, can create sharp junctions at component unions and require local smoothing or chamfering. These issues motivate an explicit, curvature-continuous parameterization with tunable smoothness that remains computationally efficient and manufacturable out of the box.

Among these methods, the implicit approaches based on pixel- or element-wise material distribution remain the most prevalent due to their generality and ease of implementation. However, topologies generated by such methods often pose challenges when transferred to manufacturing, since the resulting boundaries are jagged and contain gray regions. Such non–manufacture-ready methods typically generate structures with implicit boundary descriptions, which during post-processing require boundary extraction or geometric reconstruction. To address these limitations, extensive research efforts have been devoted by many scholars. For example, Hsu et al. \citep{hsu2005interpreting} proposed an automated interpretation framework for three-dimensional topology optimization results, in which density contour approach, smoothing, and computer-aided design (CAD) reconstruction are employed to transform gray-scale finite element outputs into smooth CAD geometries. Yi and Kim et al. \citep{yi2017identifying} developed a method that extracts implicit boundaries from topology optimization results and reconstructs CAD-ready geometries using active contours and basic CAD parametric features. Liu et al. \citep{liu2018realization} proposed an automatic realization method that transforms density-based topology optimization results into manufacturable additive manufacturing structures by skeleton extraction, boundary refinement, and adaptive B-spline fitting. In addition to the SIMP method, implicit approaches such as the Level Set method enhanced by a reaction--diffusion-based boundary regularization scheme~\citep{zhuang2021reaction}, and the Bi-directional Evolutionary Structural Optimization (BESO) method \citep{zhuang2022body}, improved through polynomial interpolation and smoothing strategies, have also been extended with corresponding techniques to refine boundary resolution and enhance manufacturability.

In contrast to these implicit formulations, explicit approaches avoid the need for boundary extraction or geometric reconstruction by directly parameterizing structural boundaries. De Ruiter and Van Keulen et al. \citep{de2000topology,de2004topology} first introduced the concept of topology description functions (TDFs) to achieve such explicit representations, laying the foundation for later developments. Among them, the Moving Morphable Components (MMC) method has emerged as a representative framework to address such challenges~\citep{guo2014doing,zhang2016new,du2022efficient}. Under the MMC framework, structural components with explicit geometry descriptions are adopted as the basic building blocks for optimization. The resulting structures produced by MMC method usually have smooth parametric boundaries, which helps to avoid the zigzag boundary issue. However, the merging and overlapping of components often lead to sharp corners or discontinuous boundaries in the transition regions. In addition, curved features are difficult to represent directly, as they are typically approximated by piecewise linear boundaries. Consequently, although the MMC framework avoids the need for feature extraction or boundary reconstruction, the resulting layouts still require additional smoothing or chamfering. Such post-processing may compromise structural performance or violate volume constraints, making the method less than ideal in practice.

Some further extensions have been developed to improve MMC method's ability to represent complex geometries and handle discontinuous connections. For instance, MMC with curved skeletons enhanced the description of curved boundaries, but sharp discontinuities at component junctions still persisted \citep{guo2016explicit}. The Moving Morphable Voids (MMV) approach employs NURBS-based void boundaries, where the adjustment and blending of control points ensure continuous curvature at connections, thereby achieving smooth transitions and maintaining CAD compatibility\citep{zhang2017explicit}. More recently, the Joint‐driven MMC (JMMC) framework incorporated additional joint components to constrain and coordinate ordinary components, thereby alleviating the “dirty geometry” issue caused by incomplete fusion
\citep{xu2025explicit}. The nonparametric geometry patching technique addresses sharp or discontinuous MMC connections by inserting spline-based patches at joints, where auxiliary control points are optimized to smooth local transitions\citep{zhang2024non}. While these methods have demonstrated notable improvements, they generally rely on inserting extra components at connections or adjusting local parametric patches, which leads to increased algorithmic complexity and stability issues.

In summary, to fully leverage the capabilities of advanced manufacturing, an ideal topology optimization framework should satisfy the following key requirements:
\begin{itemize}
    \item \textbf{Expressive geometric representation}: The framework should be able to capture complex structural features in a robust fashion, encompassing fundamental forms such as slender beams or strut members, curved or arched segments, plates, and shells within a unified modeling scheme.
    \item \textbf{Efficient optimization}: The solution pipeline should be computationally efficient, maintaining tractable solving time for a wide range of design problems and engineering contexts.
    \item \textbf{Seamless manufacturability integration}: The optimized designs should have inherent compatibility with modern manufacturing processes, without requiring extensive post-processing steps like mesh smoothing, manual feature extraction, and corner smoothing. This seamless integration offers critical advantages:  
    \begin{enumerate}
        \item By producing directly manufacturable geometries without post-processing, the method reduces intermediate steps between design and production, saving human efforts, removing subjective engineering judgments that break systematic design automation, and accelerating the product development cycle.
        \item Traditional 'optimize-then-postprocess' workflows alter the design during postprocessing in ways unaccounted for during the original optimization. These changes may violate the original optimization assumptions, potentially break the optimization constraints, and degrade performance metrics. Ideally the geometric representation should maintain consistency by natively satisfying manufacturability constraints within the optimization loop, ensuring the final output is both theoretically optimal and production-ready without post-hoc compromises.
    \end{enumerate}
\end{itemize}

Building on these motivations, we propose \textit{Gaussian Ensemble Topology (GET)}, a novel topology optimization framework that meets all of the above requirements, by representing structures using the superposition of Gaussians within a level set formulation~\cite{osher1988fronts,osher2004level}. 

In fact, superposition of Gaussians for complex function representation has been established in several computational domains. In machine learning, radial basis function networks with Gaussian kernels have served as universal approximators for scattered data interpolation \citep{scholkopf1997comparing}. The computer graphics community has adopted Gaussian splatting techniques for efficient surface rendering, where the 3D scene is represented through Gaussian basis functions \citep{kerbl20233d}. In the context of structural optimization, early works on TDFs also employed Gaussian functions \citep{de2004topology}. However, these formulations exhibited severe limitations: (i) the use of isotropic Gaussians restricted the representable geometries essentially to circular features; and (ii) the numerical implementation treated only the superposed Gaussian field heights as design variables, without allowing the Gaussians themselves to deform or move; (iii) the optimization process relied on finite-difference sensitivity analysis rather than exact analytical derivatives, which limited computational efficiency and accuracy. Consequently, the approach did not evolve further in subsequent developments.

Our proposed Gaussian Ensemble Topology combines the geometric interpretability of explicit representations with the flexibility of implicit methods, enabling smooth, high-fidelity designs that are inherently manufacturable. The superposition of simple Gaussian functions makes it computationally efficient; as discussed in Section~\ref{s41}, GET consistently outperforms MMC method in terms of iteration time under comparable settings. The key contributions of this paper are as follows:
\begin{itemize}
\item Explicit Gaussian parameterization. We represent geometry as a superposition of anisotropic Gaussian basis functions, yielding curvature-continuous boundaries without feature extraction or post-processing.

\item Efficient analytic pipeline. We derive closed-form sensitivities for all Gaussian parameters and couple them with a regularized Heaviside projection and standard FEA, enabling stable MMA optimization in 2D/3D.

\item Controllable smoothness and complexity. The threshold, band width, and number of Gaussian fields provide direct levers over boundary smoothness, discreteness, and topological complexity.

\item Competitive performance with improved geometry. On compliance and compliant-mechanism benchmarks, GET matches MMC objective values while producing smoother, CAD-friendly boundaries and fewer intermediate densities.

\item Mesh-independent parameterization. The explicit geometry can be projected onto arbitrary meshes without reparameterization, facilitating accuracy refinement and CAD export.
\end{itemize}

The rest of the paper is organized as follows. 
Section~\ref{SMethodology} introduces the Gaussian Ensemble Topology (GET) method, including the Gaussian‐based topology description, finite element discretization, and the derivation of analytical sensitivities. Section~\ref{SNumerical} presents a series of 2D and 3D numerical benchmarks, including compliance minimization and compliant mechanism problems, to demonstrate the effectiveness of the proposed framework. 
Section~\ref{SDiscussion} investigates the influence of key parameters—namely the number of Gaussian fields, the regularization parameter $\epsilon$, and the truncation threshold $T$—on structural complexity, discreteness, and boundary smoothness.
Finally, Section~\ref{SDiscussion} summarizes the main findings and outlines potential directions for future research.

\section{Methodology}\label{SMethodology}
In this section, anisotropic Gaussian functions are employed to parameterize the geometry, which is then embedded in the topology-optimization framework. Structural evolution is achieved by adjusting the parameters of the Gaussian functions. As an explicit geometry representation, superpositions of Gaussian fields provide smooth transitions in both two and three dimensions.

\subsection{Method Overview}

The Gaussian-ensemble topology optimization (GET) framework proceeds through a systematic sequence that integrates explicit geometry representation, numerical analysis, and gradient-based optimization. First, the design is parameterized by a collection of anisotropic Gaussian fields, each defined by its mean position, covariance (encoding orientation and axis lengths by rotation and scaling matrix), which together constitute the design vector. These Gaussian fields are superimposed to assemble the topology description function (TDF); thresholding the TDF at a prescribed cut-off yields a continuous geometric representation. To ensure stable numerical treatment, the TDF is further mapped to element-wise densities via a regularized Heaviside projection. Based on these densities, finite element analysis (FEA) is performed on a fixed background mesh to compute structural responses. Analytical design sensitivities of the compliance and volume constraint with respect to Gaussian parameters are derived by exact differentiation of the Gaussian fields and the adjoint method, where gradients are naturally localized to structural boundary regions. The Method of Moving Asymptotes (MMA) is then employed as the optimizer to update the Gaussian parameters iteratively, while degenerated Gaussians are automatically deactivated to improve efficiency. The optimization process continues until convergence is reached, as indicated by sufficiently small relative changes in compliance and volume fraction, at which point the procedure terminates with an optimized, explicitly defined design.

\subsection{Topology Description Based on Gaussian Functions}

The topology description function (TDF) represents the geometry in a discrete function, i.e. without intermediate densities. Let $\phi^{\mathrm{s}}$ denote the TDF over the design domain $D$. A binary geometry is obtained by thresholding $\phi^{\mathrm{s}}$ at a cut-off level $T$:
\begin{equation}
\begin{cases}
\phi^{\mathrm{s}}(\boldsymbol{x})>T, & \text{if } \boldsymbol{x}\in \Omega^{\mathrm{s}},\\
\phi^{\mathrm{s}}(\boldsymbol{x})=T, & \text{if } \boldsymbol{x}\in \partial\Omega^{\mathrm{s}},\\
\phi^{\mathrm{s}}(\boldsymbol{x})<T, & \text{if } \boldsymbol{x}\in D\setminus \Omega^{\mathrm{s}}.
\end{cases}
\end{equation}

\begin{figure}
  \centering
  \includegraphics[scale=.225]{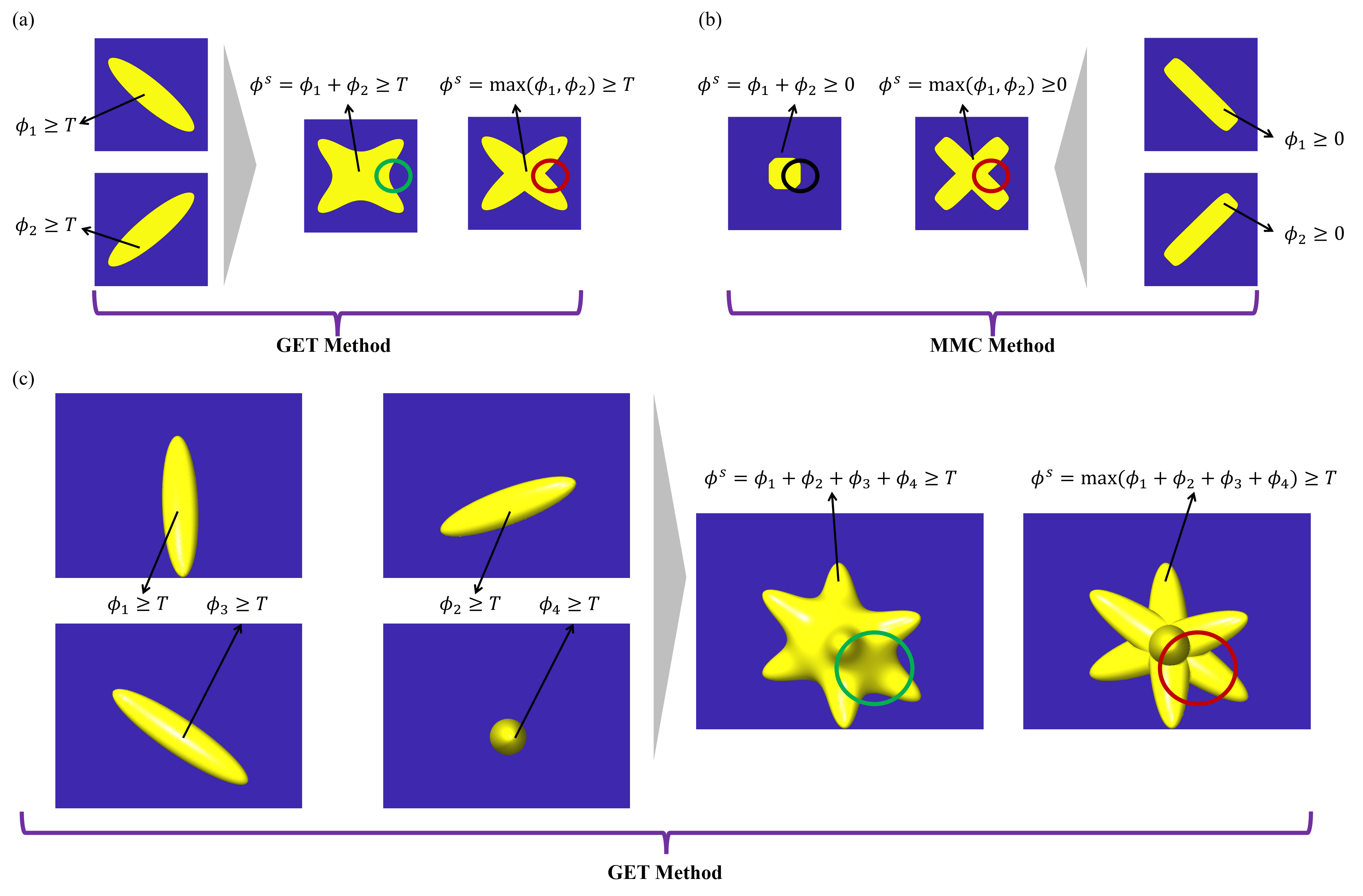}
  \caption{Geometric description and fusion in GET  and MMC Method.
  (a) 2D GET: Ellipses, represented by 2D Gaussian functions, are fused either by summation (smooth transition and union; green circle) or by the maximum (sharp transition, Boolean-like union; red circle). Solid regions are defined by $\Phi^{s} \ge T$.
  (b) 2D MMC: Super-ellipses, represented by the topology description function, are fused by summation (invalid union; black circle) or by the maximum (sharp transition, Boolean-like union; red circle). Solid regions are defined by $\Phi^{s} \ge 0$.
  (c) 3D GET: Ellipsoids, represented by 3D Gaussian functions, are fused by summation (smooth transition and union; green circle) or by the maximum (sharp transition, Boolean-like union; red circle). Solid regions are defined by $\Phi^{s} \ge T$.}
  \label{fig1}
\end{figure}

Here, $D$ is the design domain, $\Omega^{\mathrm{s}}$ is the solid region, and $\partial\Omega^{\mathrm{s}}$ is its boundary. A point $\boldsymbol{x}\in D$ is assigned material if $\phi^{\mathrm{s}}(\boldsymbol{x})>T$; otherwise, it is void, as illustrated in Fig.~\ref{fig1}.

Following this idea, we model the TDF function as a set of 2D/3D Gaussians that do not require normals. Our $i-th$ Gaussian is defined by a full covariance matrix $\Sigma_i$ centered at point (mean $\mu_i$) :

\begin{equation}
\phi_i(\boldsymbol{x})= e^ {-\frac{1}{2}(\boldsymbol{x}-\boldsymbol{\mu})^\top \boldsymbol{\Sigma}^{-1}(\boldsymbol{x}-\boldsymbol{\mu})},
\end{equation}

where $\boldsymbol{x}$ is any spatial coordinates  in the design domain and the numerical range of $\phi_i(\boldsymbol{x})$ is $[0,1]$. After truncation, each $\phi_i$ describes an ellipsoidal Gaussian field, with $\boldsymbol{\mu}_i$ representing the field’s center (with 2 parameters, $\mu_x, \mu_y$, in 2D, and 3 parameters, $\mu_x, \mu_y, \mu_z$, in 3D). The covariance matrix $\boldsymbol{\Sigma}_i$ is decomposed into a rotation matrix $\boldsymbol{R}$ and a scaling matrix $\boldsymbol{S}$:

\begin{equation}
\boldsymbol{\Sigma}=\boldsymbol{R S S}^\top \boldsymbol{R}^\top,
\end{equation}

Here, from a geometric representation perspective , $\boldsymbol{R}$ controls the Gaussian field’s orientation and $\boldsymbol{S}$ is used to control lengths of the ellipsoid’s principal axes.

In 2D, $\mathbf{R}_{2D}$ is parameterized by a single rotation angle $\theta$, and $\mathbf{S}$ is defined by two standard deviations $(\sigma_x,\sigma_y)$:
\begin{equation}
\mathbf{R}_{2D}=\begin{bmatrix}
\cos \theta & -\sin \theta \\
\sin \theta & \cos \theta
\end{bmatrix}, \qquad
\mathbf{S}_{2D}=\begin{bmatrix}
\sigma_x & 0 \\
0 & \sigma_y
\end{bmatrix}.
\end{equation}

In 3D, $\mathbf{R}_{3D}$ is parameterized by three Euler angles $(\alpha,\beta,\gamma)$, and $\mathbf{S}$ is defined by three standard deviations $(\sigma_x,\sigma_y,\sigma_z)$:
\begin{equation}
\begin{aligned}
\mathbf{R}_{3D} &= \begin{bmatrix}
\cos \beta \cos \gamma & \cos \beta \sin \gamma & -\sin \beta \\
\sin \alpha \sin \beta \cos \gamma - \cos \alpha \sin \gamma &
\sin \alpha \sin \beta \sin \gamma + \cos \alpha \cos \gamma &
\sin \alpha \cos \beta \\
\cos \alpha \sin \beta \cos \gamma + \sin \alpha \sin \gamma &
\cos \alpha \sin \beta \sin \gamma - \sin \alpha \cos \gamma &
\cos \alpha \cos \beta
\end{bmatrix}, \\
\mathbf{S}_{3D} &= \begin{bmatrix}
\sigma_x & 0 & 0 \\
0 & \sigma_y & 0 \\
0 & 0 & \sigma_z
\end{bmatrix}.
\end{aligned}
\end{equation}

\begin{figure}[htb]
  \centering
  \includegraphics[scale=.4]{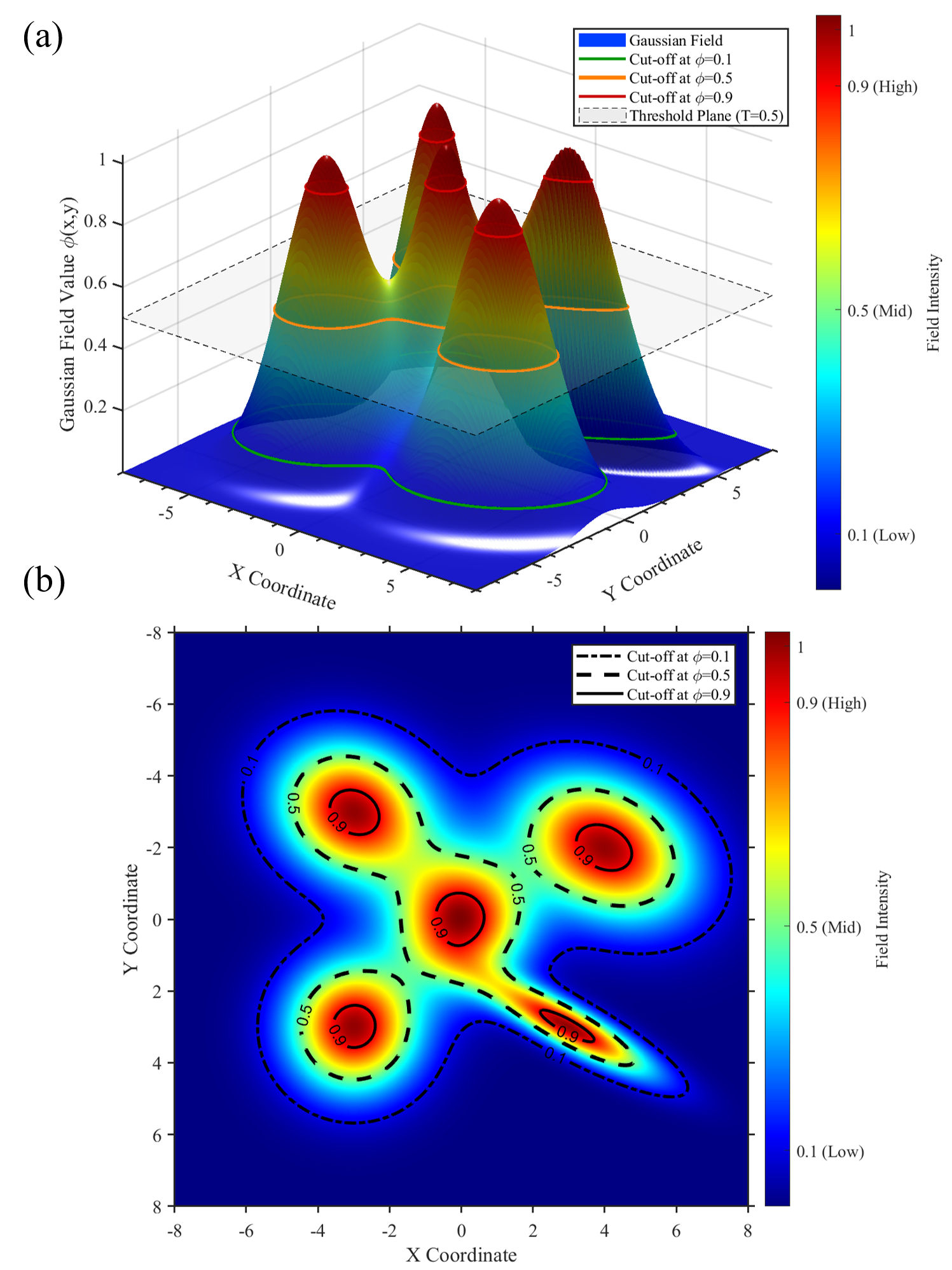}
  \caption{(a) Surface plot (height map) of five 2D Gaussian fields. 
  Thresholding at $\phi=0.1,0.5,0.9$ partitions the domain into solid ($\phi \ge T$) and void, thereby defining the geometry. 
  (b) Contour (level-set) maps at the same thresholds showing the resulting structural outlines.}
  \label{fig2}
\end{figure}

In this work, $\phi^s$ is constructed by the superposition of $n$ Gaussian fields:
$\phi^s = \sum_{i=1}^{n} \phi_i$.
Using an intuitive 2D example for further discussion, as illustrated in Fig.~\ref{fig2}, 
the composite Gaussian field formed by five superimposed components generates distinct topologies under different cut-off values $T$.
Given that the intensity of each individual Gaussian field is bounded within $[0,1]$. Consequently, the truncated geometries vary markedly with $T$. 
For instance, when $T$ is set to a high value (e.g., $0.9$), the five Gaussian fields become completely isolated, occupying only minimal regions; 
conversely, with a low $T$ value (e.g., $0.1$), the resulting structural domain expands substantially.

During the validation of numerical examples, we adopt $T=0.5$ as a balanced cut-off threshold. 
However, different choices of $T$ inherently influence the smoothness of the optimization results. 
A detailed discussion on the effects of varying $T$ is presented in Section~\ref{s5}.

This analysis highlights a key advantage of Gaussian-based geometry representation: 
the superposition of Gaussian fields inherently generates smooth and well-defined structural connections. 
As illustrated in Fig.~\ref{fig1}(a), we compare two merging approaches during Gaussian field fusion—summation versus maximum. 
When combining two Gaussian fields, the summation method preserves the original domains of each field and further creates smooth transitional regions at their interface (green circles). 
In contrast, the maximum operation produces sharp junctions (red circles) that simply represent the set union of the fields.

This intrinsic property distinguishes our method from classical explicit topology-optimization approaches. 
The Moving Morphable Component (MMC) method, a milestone in explicit TO, exhibits related limitations. 
Since MMC employs signed-distance-type $\phi$-fields (e.g., ranging in $(-\infty,1]$) to represent components, where $\phi\ge 0$ represents solid region, it uses maximum operations for merging, which inevitably creates sharp connections. 
Moreover, due to the rapid decay of field values away from components, summation in MMC tends to approximate intersection rather than union (black circles in Fig.~\ref{fig1}(b).

The proposed method achieves smooth geometric transitions while maintaining explicit representation, and this advantage becomes even more pronounced in 3D. 
As shown in Fig.~\ref{fig1}(c), the superposition of four 3D Gaussian functions produces seamless transitional surfaces(green circles), whereas the maximum operation yields unnaturally sharp edges at all junctions(red circles).

The automatic generation of smooth transitions is significant for structural design:
\begin{itemize}
  \item \textbf{Geometric representation:} Sharp junctions based on MMC method typically require multiple components to approximate complex curvature (i.e., piecewise-linear approximations), whereas Gaussian superposition naturally achieves curvature continuity. Moreover, for the structural backbone, suitable superpositions of Gaussian fields can also produce long, nearly straight members(see Section~4 for detailed examples), demonstrating strong geometric expressiveness. 
  \item \textbf{Mechanical performance:} Curvature-continuous and smooth transitions remove sharp corners and abrupt changes, thereby lowering local stress concentrations and steep strain gradients. This intrinsic design features can lower peak stresses, increases static strength margins, and delays fatigue-crack initiation.
  \item \textbf{Manufacturability and post-processing:} Topology-optimized structures often require complex post-processing to address manufacturing defects and stress concentrations (e.g., additional smoothing and fillets at transitions).   In contrast, the Gaussian-based representation provides an explicit, mesh-independent vector geometry amenable to CAD, and it inherently yields sufficiently smooth topologies with naturally filleted junctions—thereby reducing or eliminating traditional post-processing while maintaining mechanical performance and volume constraints.
\end{itemize}

In summary, under the above geometry-representation scheme, the layout (i.e., shape and topology) of a structure is fully determined by a design vector
$\boldsymbol{D} = (\boldsymbol{D}_1, \boldsymbol{D}_2, \ldots, \boldsymbol{D}_n)$,
where $\boldsymbol{D}_i = (\mu_{xi}, \mu_{yi}, \sigma_{xi}, \sigma_{yi}, \theta_i)$ in 2D (five design parameters) and 
$\boldsymbol{D}_i = (\mu_{xi}, \mu_{yi}, \mu_{zi}, \sigma_{xi}, \sigma_{yi}, \sigma_{zi}, \alpha_i, \beta_i, \gamma_i)$ in 3D (nine design parameters).

\subsection{Problem formulation}
The topology optimization problem under the Gaussian-function-based framework can be formulated as:
\begin{equation}
\begin{aligned}
\text{Find}\quad & \boldsymbol{D}=\big(\left(\boldsymbol{D}^1\right)^{\top},\ldots,\left(\boldsymbol{D}^n\right)^{\top}\big)^{\top},\\
\text{Minimize}\quad & I=I(\boldsymbol{D}),\\
\text{s.t.}\quad & g_j(\boldsymbol{D}) \le 0,\quad j=1,\ldots,m,\\
& \boldsymbol{D} \in \mathcal{U}_D.
\end{aligned}
\label{eq6}
\end{equation}

For minimum-compliance design with a finite element discretization, the problem reads:
\begin{equation}
\begin{aligned}
\text{Find}\quad & \boldsymbol{D}=\big(\left(\boldsymbol{D}^1\right)^{\top},\ldots,\left(\boldsymbol{D}^n\right)^{\top}\big)^{\top},\\
\text{Minimize}\quad & C=\boldsymbol{f}^{\top}\boldsymbol{u},\\
\text{s.t.}\quad & \boldsymbol{K}\,\boldsymbol{u}=\boldsymbol{f},\\
& g = V/|\mathrm{D}| - \bar{v} \le 0,\\
& \boldsymbol{u}=\overline{\boldsymbol{u}} \ \text{on}\ \Gamma_{\mathrm{u}},\\
& \boldsymbol{D} \in \mathcal{U}_D.
\end{aligned}
\label{eq:mincomp}
\end{equation}
where $\boldsymbol{K}$, $\boldsymbol{F}$, and $\boldsymbol{u}$ denote the global stiffness matrix, the nodal force vector, and the nodal displacement vector, respectively; $V$, $|\mathrm{D}|$, and $\bar{v}$ denote the volume of the optimized structure, the volume of the design domain, and the upper bound on the allowable volume fraction; and $\mathcal{U}_D$ is the admissible set of design variables.

\subsection{Numerical implementation}

To solve the above optimization problem, the structural response $\boldsymbol{u}$ is obtained via finite element analysis (FEA). Since geometry boundaries are explicitly described within the Gaussian framework, the Gaussian field can be projected onto an arbitrary mesh on the fly. Thus, either a fixed background mesh or an adaptively updated mesh can be employed. This decoupling from the background mesh is a key advantage of explicit optimization algorithms. In this study, to enhance computational efficiency, we consistently use a fixed background mesh.

Specifically, the global stiffness matrix is assembled from the elemental stiffness of the $e$-th element,
\begin{equation}
    \boldsymbol{k}_e \;=\; \rho_e\,\boldsymbol{k}^0
    \;=\; \frac{\boldsymbol{k}^0}{N_{NpE}}\sum_{p=1}^{N_{NpE}} H_\epsilon^\alpha\!\left(\phi^{\mathrm{s}}_{ep}\right),
    \label{eq7}
\end{equation}
where $\rho_e$ and $\boldsymbol{k}^0$ are the density and the elemental stiffness matrix of the base material for the $e$-th element, respectively. The symbol $N_{NpE}$ denotes the number of nodes per element—four-node bilinear quadrilateral elements for 2D analysis ($N_{NpE}=4$) and eight-node trilinear hexahedral elements for 3D analysis ($N_{NpE}=8$). The value $\phi^{\mathrm{s}}_{ep}$ is the TDF evaluated at node $p$ of element $e$. The regularized Heaviside function $H_\epsilon^\alpha$ is given by
\begin{equation}
H(\phi)=
\begin{cases}
1, & \text{if } \phi > T+\epsilon,\\[2pt]
\alpha, & \text{if } \phi < T-\epsilon,\\[2pt]
\dfrac{3(1-\alpha)}{4}\!\left(\dfrac{\phi-T}{\epsilon}-\dfrac{(\phi-T)^3}{3\,\epsilon^3}\right)
+\dfrac{1+\alpha}{2}, & \text{otherwise,}
\end{cases}
\label{eq9}
\end{equation}
where $\epsilon$ is the width of the transition band, $T=0.5$ is the cut-off value discussed in the previous section, and $\alpha=10^{-3}$ is a small positive number introduced to avoid the possible singularity of the 
global stiffness matrix associated with a void design. Once the Young’s modulus and corresponding density of each element is determined, FEA can be performed straightforwardly.

\begin{figure}[htb]
  \centering
  \includegraphics[scale=.38]{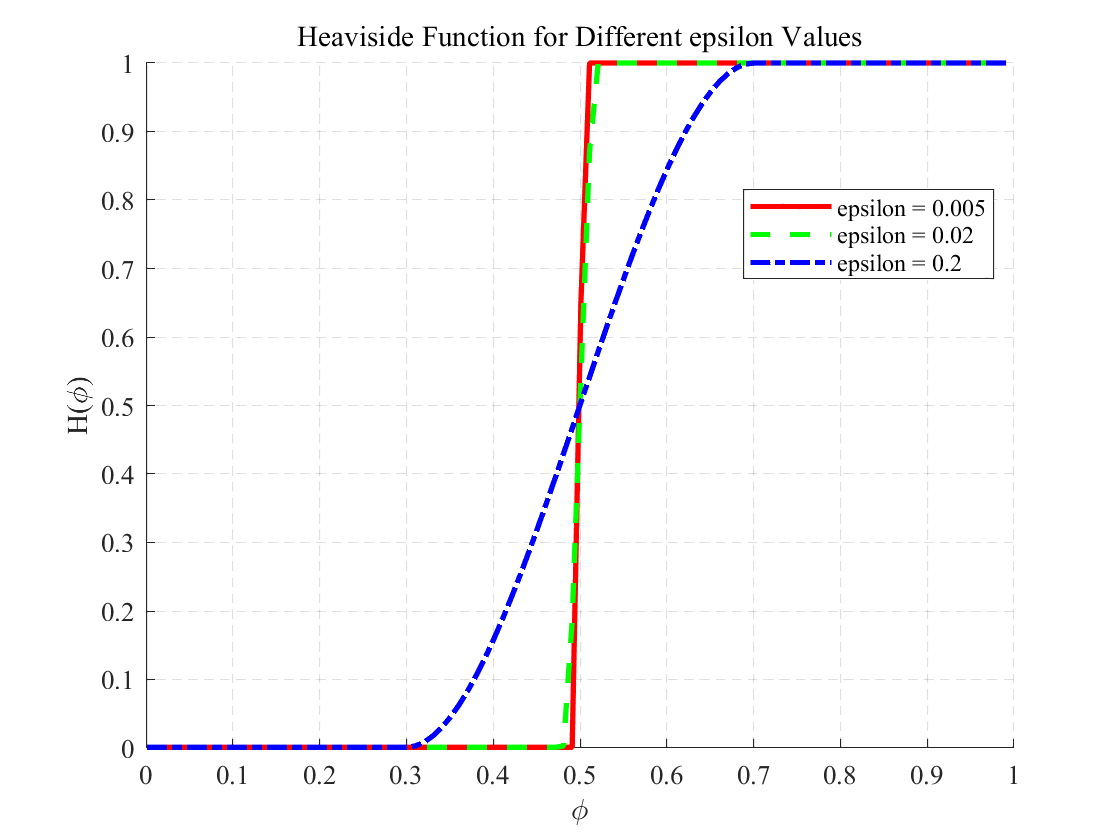}
  \caption{Density–phase distribution of the Gaussian field after applying the Heaviside function with $\epsilon\in\{0.005,\,0.02,\,0.2\}$, illustrating how $\epsilon$ controls the extent of intermediate densities.}
  \label{fig3}
\end{figure}

Figure~\ref{fig3} shows the density–phase distribution obtained for different $\epsilon$ values. After superposition, $\phi^{\mathrm{s}}$ typically lies in $[0,1+]$. The Heaviside mapping drives most element densities toward $0$ or $1$, while $\epsilon$ sets the width of the transition band—the smaller $\epsilon$, the sharper the black–white separation; after optimization, setting $\epsilon$ to zero yields a strictly binary (0–1) design. Moreover, its S-shaped profile increases the sensitivity of cut elements at intermediate densities and thereby promotes the evolution of structural boundaries.

In the numerical examples of this study, we set $\epsilon=0.02$ unless otherwise specified. Compared with the MMC method, where $\epsilon$ is often chosen in the range $0.1$–$0.2$, the Gaussian-field representation permits a smaller $\epsilon$ due to its concentrated superposition, yielding fewer intermediate densities and clearer boundaries.

During the optimization, to accelerate convergence and improve efficiency, any Gaussian field is deactivated (i.e., stops contributing intensity and sensitivity) when either $\sigma_x$ or $\sigma_y$ drops below one-half of the element edge length.

\subsection{Design sensitivity}

Only the design sensitivity for the compliance-minimization problem is discussed here. Extensions to more general objective/constraint functions can be obtained via the adjoint method.

If the objective is the structural compliance, its sensitivity with respect to an arbitrary design variable \(d\) is
\begin{equation}
\frac{\partial C}{\partial d}
= \frac{\partial C}{\partial \rho_e}\frac{\partial \rho_e}{\partial d}=  -\,\boldsymbol{u}^{\top}\frac{\partial \boldsymbol{K}}{\partial \rho_e}\,\boldsymbol{u}\frac{\partial \rho_e}{\partial d},
\end{equation}
where the global stiffness only depends on the element densities,
\(\boldsymbol{K}=\sum_{e=1}^{N_{Ele}}\rho_e\,\boldsymbol{k}^0\) with
\(\rho_e=\tfrac{1}{N_{NpE}}\sum_{p=1}^{N_{NpE}} H_\epsilon^\alpha\!\left(\phi_{ep}^{\mathrm{s}}\right)\).
Therefore,
\begin{equation}
\frac{\partial C}{\partial d}
= - \sum_{e=1}^{N_{Ele}}
\Bigg(\frac{\partial \rho_e}{\partial d}\,\boldsymbol{u}^{\top}\boldsymbol{k}^0\boldsymbol{u}\Bigg),
\qquad
\frac{\partial \rho_e}{\partial d}
= \frac{1}{N_{NpE}}\sum_{p=1}^{N_{NpE}}
\frac{\partial H_\epsilon^\alpha\!\left(\phi_{ep}^{\mathrm{s}}\right)}{\partial \phi_{ep}^{\mathrm{s}}}\,
\frac{\partial \phi_{ep}^{\mathrm{s}}}{\partial d}.
\end{equation}

Similarly, the volume sensitivity is computed consistently as
\begin{equation}
\frac{\partial V}{\partial d}
= \frac{1}{N_{NpE}}\sum_{e=1}^{N_{Ele}}\sum_{p=1}^{N_{NpE}}
\frac{\partial H_\epsilon^\alpha\!\left(\phi_{ep}^{\mathrm{s}}\right)}{\partial \phi_{ep}^{\mathrm{s}}}\,
\frac{\partial \phi_{ep}^{\mathrm{s}}}{\partial d}.
\end{equation}

Since the Gaussian functions are superposed by summation, \(\phi^{\mathrm{s}}=\sum_{i}\phi_i\), the chain rule for the \(m\)-th parameter of the \(i\)-th component gives
\begin{equation}
\frac{\partial H_\epsilon^\alpha\!\left(\phi^{\mathrm{s}}\right)}{\partial d_{im}}
= \frac{\partial H_\epsilon^\alpha\!\left(\phi^{\mathrm{s}}\right)}{\partial \phi^{\mathrm{s}}}\,
\frac{\partial \phi^{\mathrm{s}}}{\partial d_{im}}
= \frac{\partial H_\epsilon^\alpha\!\left(\phi^{\mathrm{s}}\right)}{\partial \phi^{\mathrm{s}}}\,
\frac{\partial \phi^{\mathrm{s}}}{\partial \phi_{i}} \frac{\partial \phi_{i}}{\partial d_{im}}= 
\frac{\partial H_\epsilon^\alpha\!\left(\phi^{\mathrm{s}}\right)}{\partial \phi^{\mathrm{s}}}\,
\frac{\partial \phi_{i}}{\partial d_{im}},
\end{equation}

where \(\partial \phi^{\mathrm{s}}/\partial \phi_i = 1\).  $\partial \phi_i/\partial d_{im}$ follows by exact differentiation (see Appendix for closed forms). For second term $\frac{\partial \phi_{i}}{\partial d_{im}}$, it means that each design variable's sensitivity is only related to its own Gaussian component; For the first term $\frac{\partial H_\epsilon^\alpha\!\left(\phi^{\mathrm{s}}\right)}{\partial \phi^{\mathrm{s}}}$, due to the Heaviside projection, it is only related to oneself and neighborhood.
consequently, analysis process with respect to parameters of different Gaussian functions are decoupled. This property enables straightforward parallel implementation and improves scalability for large-scale designs.

The derivative of the regularized Heaviside is
\begin{equation}
\frac{\partial H_\epsilon^\alpha(\phi^{\mathrm{s}})}{\partial \phi^{\mathrm{s}}}
=
\begin{cases}
0, & \text{if } \phi^{\mathrm{s}} > T+\epsilon,\\[2pt]
0, & \text{if } \phi^{\mathrm{s}} < T-\epsilon,\\[2pt]
\dfrac{3(1-\alpha)}{4\,\epsilon}\!\left(1-\dfrac{(\phi^{\mathrm{s}}-T)^2}{\epsilon^{2}}\right), & \text{otherwise.}
\end{cases}
\end{equation}

Note that $\dfrac{\partial H_\epsilon^\alpha(\phi^{\mathrm{s}})}{\partial \phi^{\mathrm{s}}}=0$ outside the band $[T-\epsilon,\,T+\epsilon]$; hence only cut elements near the boundary contribute to the gradients, which localizes updates and improves numerical stability. In particular, weak-density (void-region) elements with $\rho_e=\alpha$ do not contribute to the sensitivities. Although we retain all DOFs in this work for consistency, this observation indicates a potential acceleration for future implementations— excluding such elements in corresponding iteration to reduce the size of DoFs. and improve computational efficiency.

\section{Numerical examples}\label{SNumerical}

In this section, several representative 2D and 3D examples are presented to demonstrate the effectiveness of the proposed approach. Since only numerical performance is of interest, material properties, and external loads are treated as dimensionless. The finite element discretizations employ four-node bilinear quadrilateral elements for 2D analysis and eight-node trilinear hexahedral elements for 3D analysis. The Young’s modulus and Poisson’s ratio of the solid material are set to $E=1$ and $\nu=0.3$, respectively. The Method of Moving Asymptotes (MMA) \citep{svanberg1987method} is used as the optimizer. All code in this work was implemented in MATLAB2024a and executed on a desktop computer equipped with a 13th-Gen Intel Core i7-13700KF CPU.

\subsection{2D numerical examples}

\begin{figure}[ht]
  \centering
  \includegraphics[scale=.35]{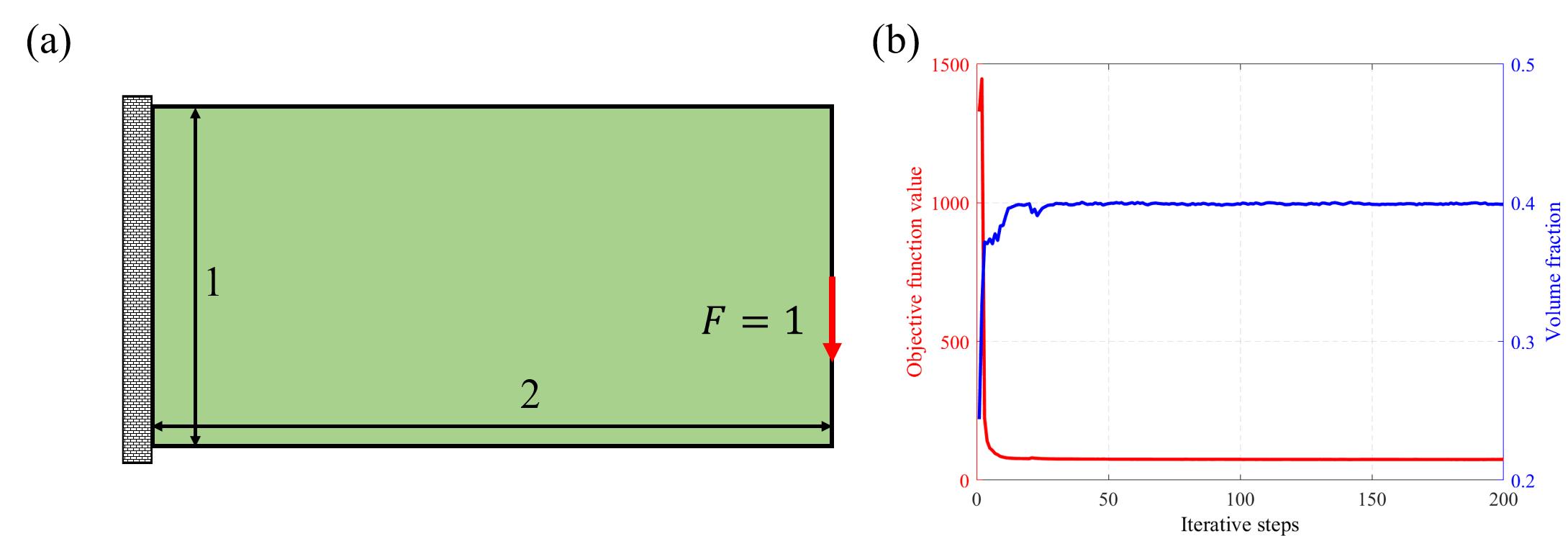}
  \caption{(a) Design domain and boundary conditions of the 2D cantilever beam; (b) Convergence history of the objective and volume fraction for the cantilever beam obtained by the GET method.}
  \label{fig4}
\end{figure}

We first investigate the 2D problems to illustrate the method’s core behavior. The study proceeds through three types of benchmarks. 
(i) A detailed cantilever beam case establishes baseline performance and is compared with the MMC method in detail. 
(ii) Other classical examples are used to verify symmetry maintenance, solid/void non-design regions and geometric representation. 
(iii) Finally, compliant mechanism problems are considered, demonstrating the framework’s versatility across distinct optimization settings.

\begin{figure}[htb]
  \centering
  \includegraphics[scale=.20]{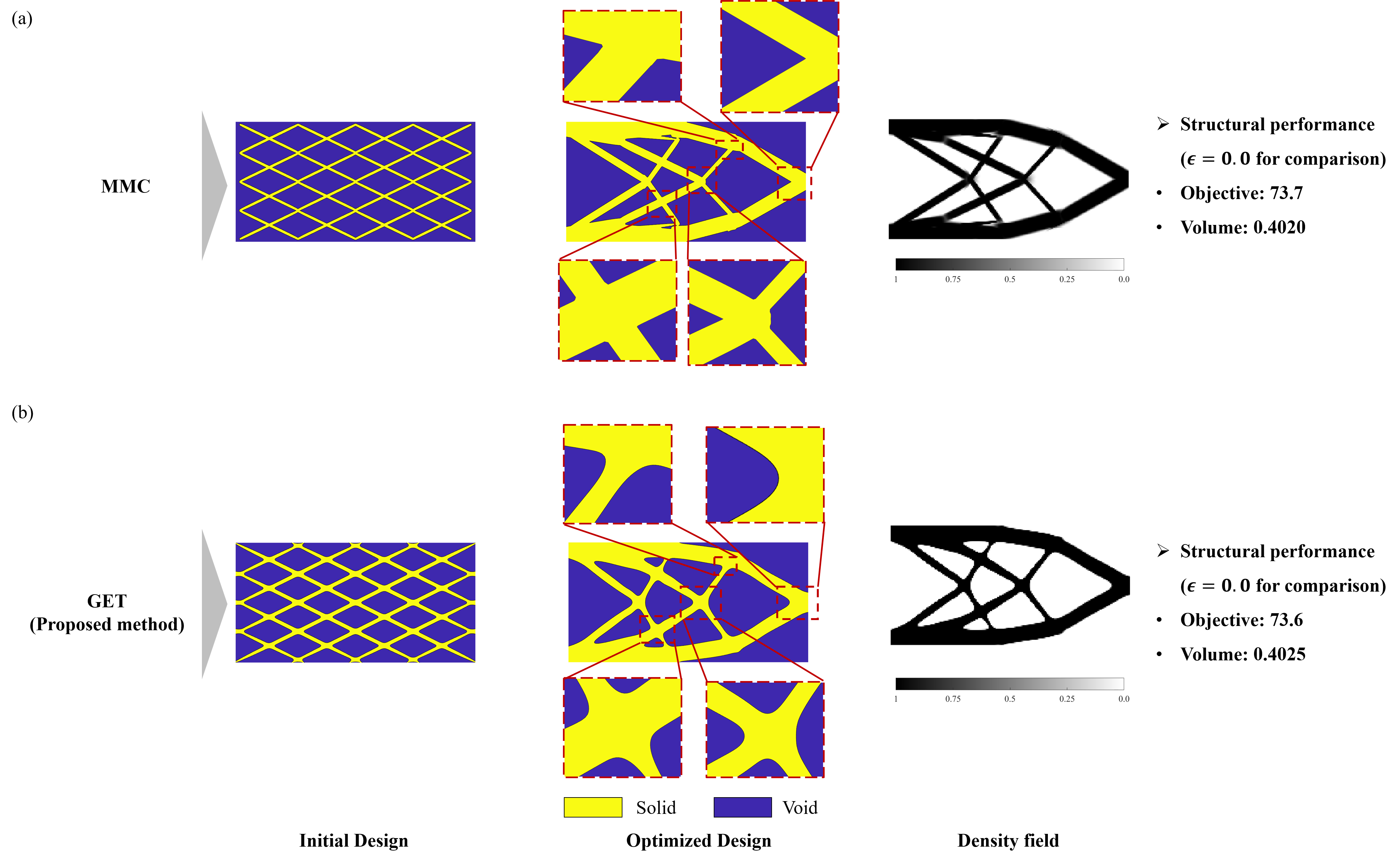}
  \caption{2D cantilever beam: (a) MMC method: initial design and optimized structure (yellow: solid; blue: void; red box: sharp connection; grayscale: density field). Final performance ($\epsilon=0$ for comparison): $C=73.7$, $V_f=0.4020$. (b) Proposed GET method: similar initial design and optimized structure (red box: smoother connections with fewer gray elements). Final performance ($\epsilon=0$ for comparison): $C=73.6$, $V_f=0.4025$.}
  \label{fig5}
\end{figure}

\subsubsection{2D Cantilever beam example}

The proposed GET method is first validated on the classical cantilever beam benchmark. The design domain, boundary conditions, and applied load are shown in Fig.~\ref{fig4}(a). The rectangular domain of length~2 and height~1 is discretized by a $200\times100$ FE mesh. A unit vertical load ($F_y=1$) is applied at the midpoint of the free right edge. The volume-fraction is set to $0.4$.

For comparison, both the MMC and GET methods, initialized with a similar X-shaped and staggered array of components, are used to optimize this design problem. For MMC method, we use trapezoidal variable-width components (6 design variables each, 192 design variables in total) and hyperparameters consistent with public benchmark implementations \citep{du2022efficient}. For GET method, the layout is encoded by $32=4\times4\times2$ Gaussian fields, giving $160=32\times5$ design variables in total. This number is much smaller than that in traditional implicit schemes and is independent of the FE mesh resolution.

\begin{figure}[ht]
  \centering
  \includegraphics[scale=.475]{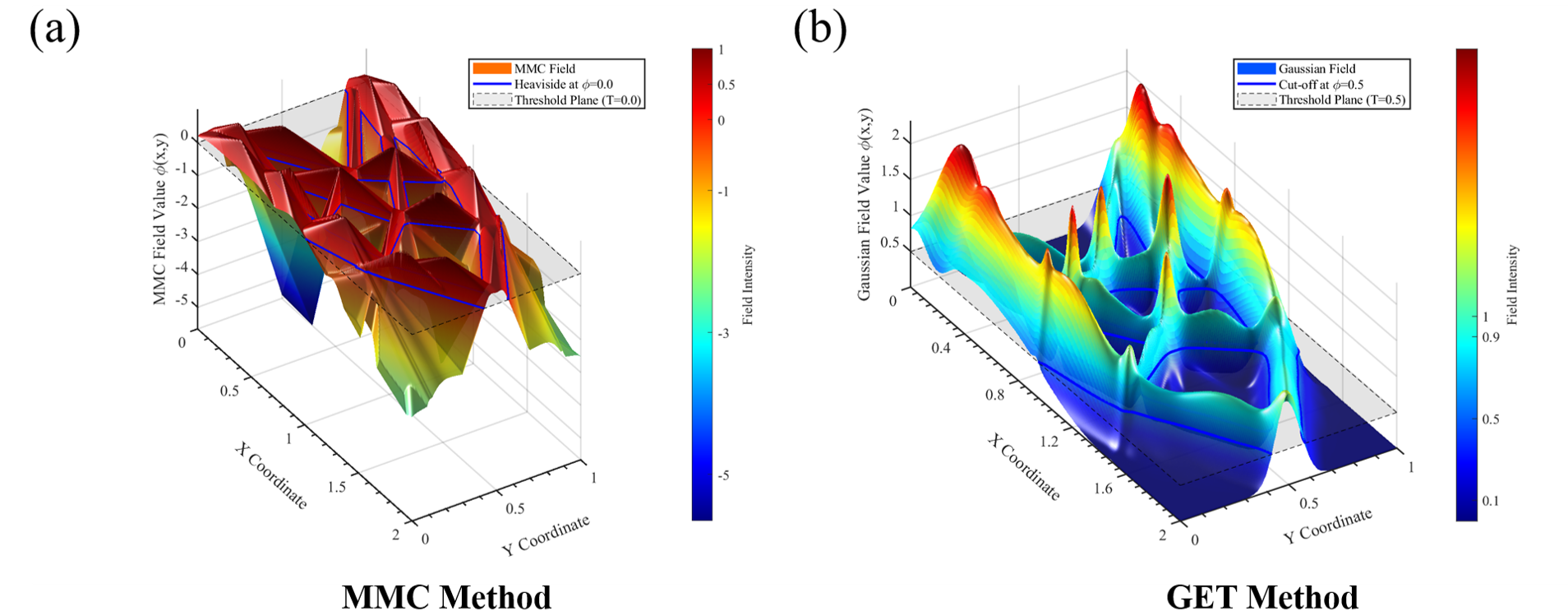}
  \caption{2D cantilever beam: (a) MMC: surface plot (height map) of the $\phi^{\mathrm{s}}$ field with $\phi^{\mathrm{s}}=\max_i(\phi_i$) and Heaviside cut-off at $0$; peak $\phi^{\mathrm{s}}$ occurs near component cores. (b) GET: surface plot of the $\phi^{\mathrm{s}}$ field with $\phi^{\mathrm{s}}=\sum_i \phi_i$ and cut-off at $0.5$; peak $\phi^{\mathrm{s}}$ concentrates at Gaussian-field connections.}
  \label{fig6}
\end{figure}

The optimized structures obtained by the MMC and GET methods are shown in Fig.~\ref{fig5}. Under identical initializations, both methods produce similar topologies (yellow represents solid structure; blue represents void region) and nearly identical objective values. However, as highlighted by the red boxes in Fig.~\ref{fig5}(a), the design from the MMC method exhibits sharp corners at component junctions. Its result usually requires additional smoothing to avoid stress concentration and fabrication defects. In contrast, as shown in Fig.~\ref{fig5}(b),design from GET method naturally yields smooth transitions in these regions, avoiding further post-processing. The third panel in Fig.~\ref{fig5} further visualizes these differences on the background mesh: MMC method shows some gray elements and slight protrusions at connections, whereas GET method confines intermediate densities to a narrow band around the boundary and preserves smooth junctions—even on this display mesh with only $2\times10^4$ elements. 

The convergence histories in Fig.~\ref{fig4}(b) track the compliance $C$ (red) and volume fraction $v_f$ (blue) over 200 iterations, showing that $C$ decreases steadily to convergence while the volume constraint is satisfied. For structural performance, to ensure a fair comparison between the two final designs, we set $\epsilon=0$ to obtain strictly black-and-white structures for the FE evaluation. The final metrics are: MMC ($C=73.7$, $V_f=0.4020$) and GET ($C=73.6$, $V_f=0.4025$). GET method attains compliance comparable to MMC while delivering smoother connections. For the volume constraint, note that classical implicit SIMP schemes often produce many gray elements and require extra filtering/projection; truncation directly may violate the volume bound. In contrast, design by GET method achieves a binary layout simply by taking $\epsilon\!\to\!0$, with only a $0.625\%$ overshoot relative to the target bound. A detailed discussion of structural discreteness is provided in Section~\ref{S5.2}.

Conceptually, both the GET and MMC methods can be viewed as thresholdings of explicitly constructed level-set–like fields. Figure~\ref{fig6} provides an intuitive comparison. For GET method, $\phi^{\mathrm{s}}=\sum_i \phi_i$: the structural backbone forms “ridges,” with “peaks” at field connections establishing smooth load-transfer paths, while void regions without fields become “valleys” . For MMC, $\phi^{\mathrm{s}}=\max_i(\phi_i$) with range $(-\infty,1]$: the field attains its maximum at component centers and decays toward the component's ends; near the boundary it drops steeply, creating 'cliff'-like height map that leads to abrupt contour and curvature changes. Compared with MMC method, GET method provides smoother transitions between peaks, ridges, and valleys, thus preserving geometric smoothness in the thresholded boundaries.

\subsubsection{2D MBB beam example}
\begin{figure}[ht]
  \centering
  \includegraphics[scale=.225]{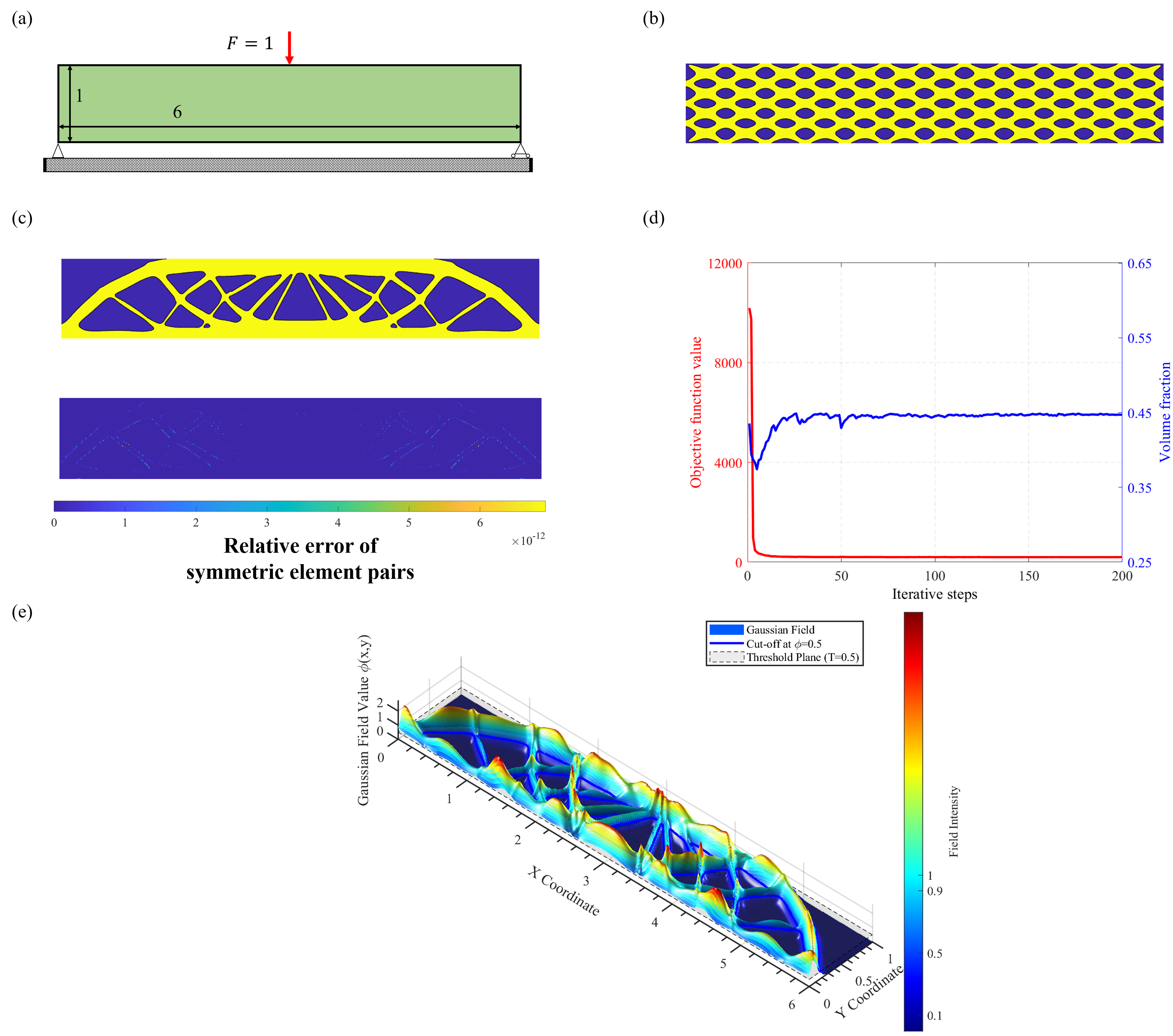}
  \caption{(a) Boundary conditions of the 2D MBB beam; (b) initial layout of Gaussian fields; (c) optimized configuration; (d) convergence history; (e) surface plot (height map) of Gaussian fields in the optimized layout.}
  \label{fig7}
\end{figure}
The MBB (Messerschmitt–Bölkow–Blohm) beam, a standard benchmark in topology optimization, is used to further assess the proposed methodology. The design domain and boundary conditions are shown in Fig.~\ref{fig7}(a). The design has dimensions $6\times1$, with the left end fixed and the right end simply supported to allow vertical displacement; a $600\times100$ FE mesh is used. A concentrated vertical load ($F_y=1$) is applied at the top-middle point, and the volume-fraction bound is $0.45$.

The initial Gaussian basis is arranged uniformly and in a symmetric pattern ($12\times4\times2=96$ fields, $96\times5=480$ design variables), as depicted in Fig.~\ref{fig7}(b). Although the MBB example's optimal design admits symmetric configuration so that one can  model only half of the domain for efficiency, we analyze the full domain here to demonstrate that without dedicated enforcement, the GET method can preserve symmetry when the problem and initial layout are symmetric. All subsequent symmetric examples will adopt symmetry-reduced models unless otherwise specified. 

After 200 iterations (see the convergence history in Fig.~\ref{fig7}(d)), the optimized configuration in Fig.~\ref{fig7}(c) is comparable to conventional results while retaining superior geometric regularity along the boundaries, especially at junctions. The surface plot of the Gaussian fields in Fig.~\ref{fig7}(e) exhibits a totally symmetric distribution. To reduce the accumulation of truncation errors and help preserve symmetry, the design sensitivities are rounded to five significant digits at each iteration. Fig.~\ref{fig7}(c) shows the relative error in the density of symmetric element pairs of the last iteration; the error is concentrated in intermediate-density elements near the boundary, with a maximum of $6.92\times10^{-12}\%$. If $\epsilon$ is set to $0$ to obtain a design with no intermediate densities, the relative error becomes $0$, i.e., the final structure is exactly symmetric. This resulting symmetric layout indicates that neither the mathematical derivation nor the discretization implementation introduces directional bias, which further validates the correctness and robustness of the proposed algorithm.

\subsubsection{L-shaped beam example}
\begin{figure}[h]
	\centering
		\includegraphics[scale=.25]{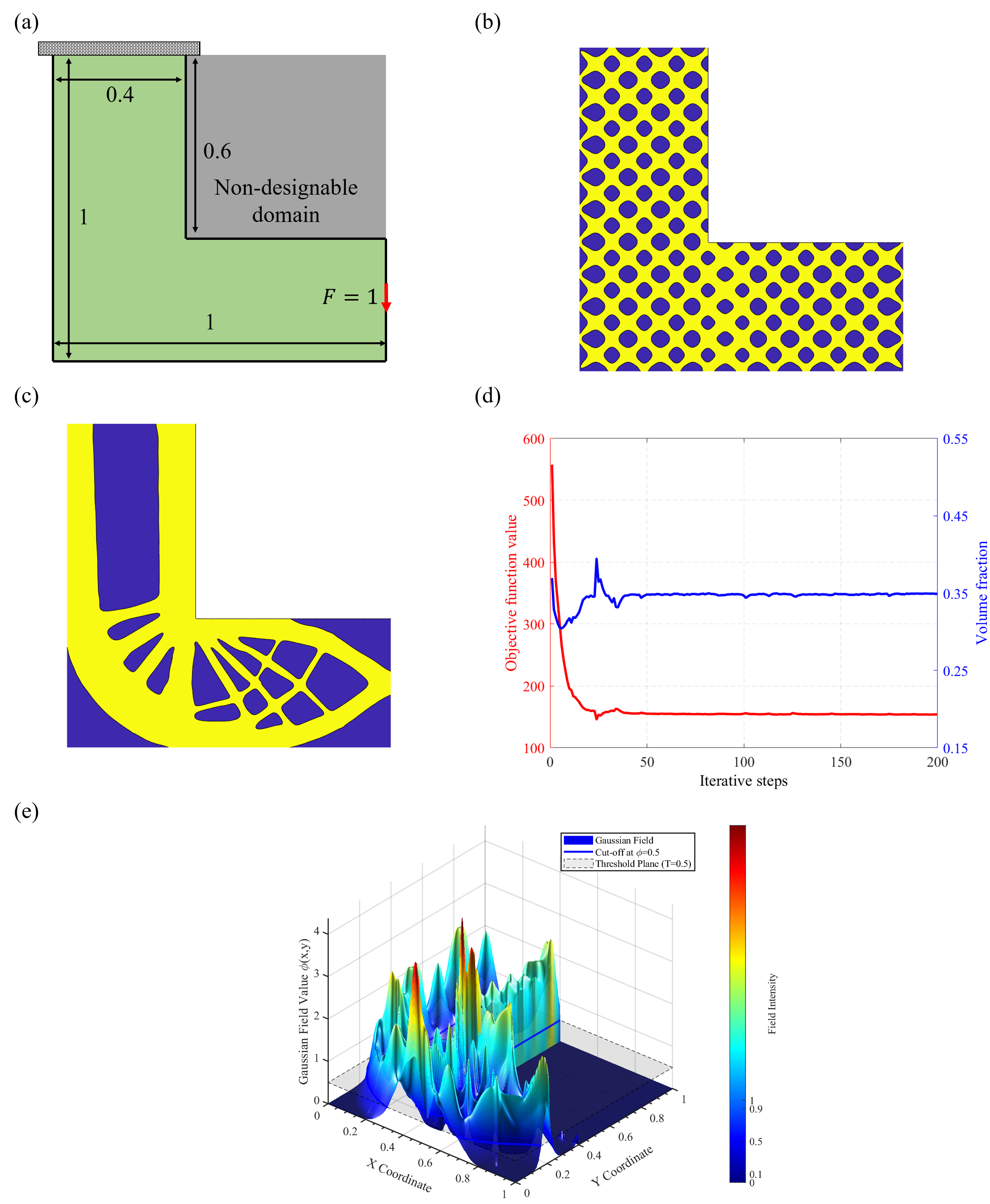}
	\caption{(a) Boundary conditions of the 2D L-shaped beam example; (b) Initial layout of Gaussian fields; (c) Optimized configuration; (d) Iterative convergence curve; (e) Surface plot (height map) of Gaussian fields in the optimized configuration}
	\label{fig8}
\end{figure}
The proposed GET method is further validated on an L-shaped beam featuring asymmetric geometry and a prescribed non-design void. The design domain and boundary conditions are shown in Fig.~\ref{fig8}(a). The $1\times1$ domain is discretized by a $200\times200$ FE mesh. A square non-design void of size $0.6\times0.6$ is placed at the top-right corner and assigned a fixed near-zero density for numerical stability ($\rho_{\text{void}}=10^{-3}$). The top-left edge is fully fixed, and a concentrated downward load ($F_y=1$) is applied at the midpoint of the right edge. The volume-fraction bound is $0.35$.

To assess robustness and numerical stability under a larger field count, this example deliberately employs a relatively large number of Gaussian fields. The initial Gaussian basis is uniformly distributed (128 Gaussian fields; 640 design variables), as shown in Fig.~\ref{fig8}(b). The optimization converges in 200 iterations; see Fig.~\ref{fig8}(d).

The optimized topology in Fig.~\ref{fig8}(c) retains the merits of GET method, exhibiting smooth, curvature-continuous boundaries while strictly respecting the non-design void (no Gaussian components are placed in the void region and the local density remains clamped at $\rho_{\text{void}}$). The GET method converges to the classical L-beam layout. Figure~\ref{fig8}(e) further shows the surface plot of the superposed Gaussian field.  Due to the increased design dimensionality from using more design parameters, a highly corrugated height map with dense peaks and ridges generated by numerous overlapping functions; nevertheless, thresholding at $T=0.5$ cleanly extracts the reasonable optimized structure.

\subsubsection{2D Bridge-like beam example}
\begin{figure}[ht]
	\centering
		\includegraphics[scale=.2]{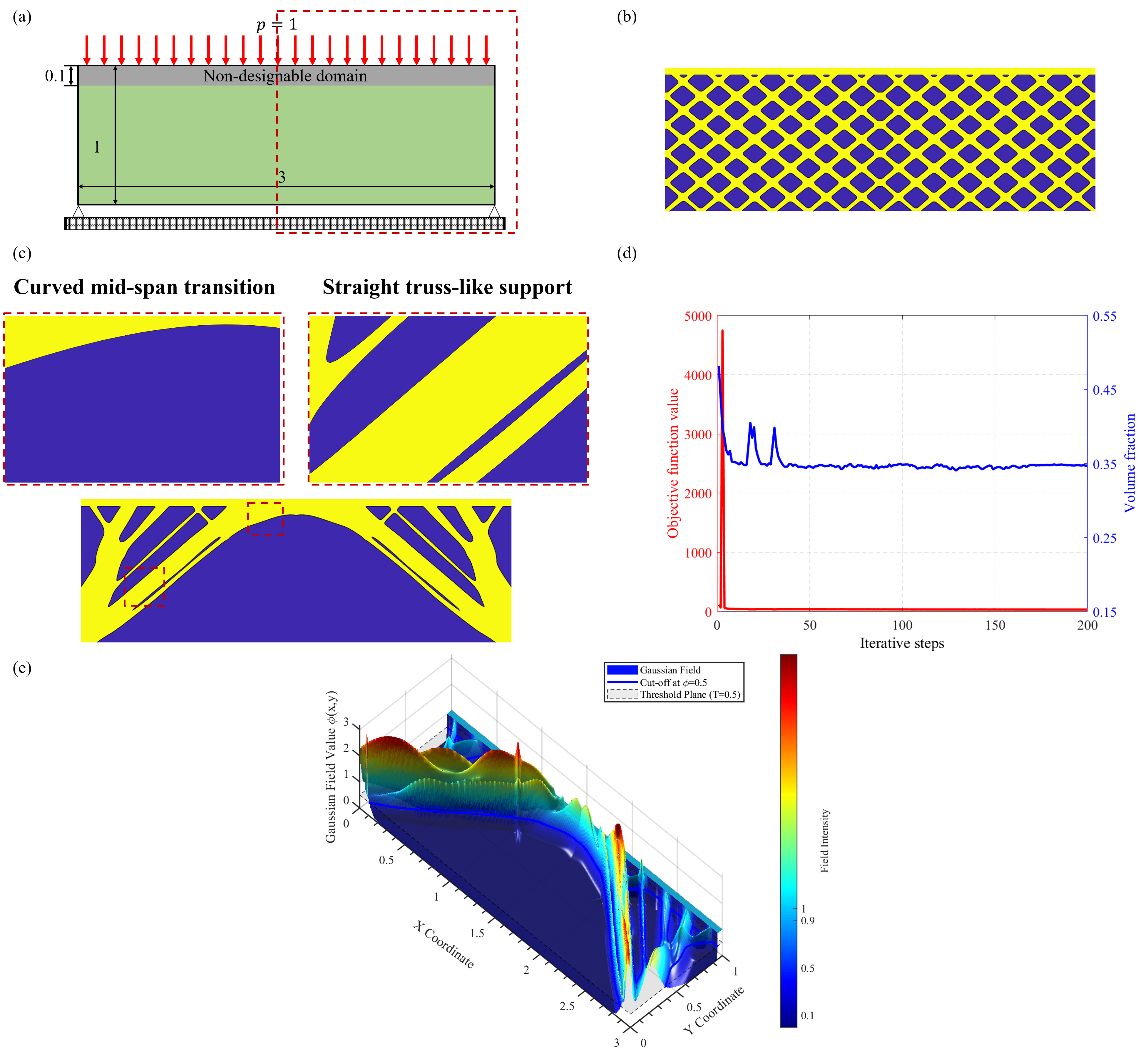}
	\caption{(a) Boundary conditions of the 2D Bridge-like beam example; (b) Initial layout of Gaussian fields; (c) Optimized configuration; (d) Iterative convergence curve; (e) Surface plot (height map) of Gaussian fields in the optimized configuration}
	\label{fig9}
\end{figure}
 In order to demonstrate the flexibility of geometry description of the proposed approach, this bridge-like beam problem whose optimal solutions may include parts with truss-like support segments and curved transition at mid-span. The design domain and boundary conditions are illustrated in Fig. \ref{fig9}(a). The bridge beam is subjected to a uniformly distributed pressure load ($p = 1$) acting vertically downward on its top surface. Pin supports constrain both the bottom-left and bottom-right corners ($u_x = 0$, $u_y = 0$). In this example, the design domain is discretized by a $300 \times 100$ FE mesh, and a rectangular zone of size $0.1 \times 3$ at the top of the domain is set as a fixed solid region. The volume fraction constraint is set to $0.35$. Under symmetric boundary conditions, only half of the structure is analyzed. The initial design after the symmetry operation, as shown in Fig. \ref{fig9}(b). During optimization, this case consists of $6\times5\times2 = 60$ Gaussian fields, corresponding to $60\times5 = 300$ design variables. The convergence curve is shown in Fig. \ref{fig9}(d) after 200 iterations.

The optimized topology and Gaussian fields surface plot (height map) are presented in Fig. \ref{fig9}(c) and (e). The optimized configuration exhibits characteristic arch-bridge supporting structures. At the mid-span, Gaussian functions' superposition generates smooth curved transitions (whereas the MMC method’s inherently rectangular components can only approximate curves through piecewise-linear connections or need more complex topology description). For the truss-like supports, while individual Gaussian fields may struggle to form straight connections - particularly when $\sigma_x$ and $\sigma_y$ differ, resulting in oblate spheroidal shapes with thicker central regions and thinner edges (most evident in the initial layout) - the final structure in Fig. \ref{fig9}(c) demonstrates that multiple overlapping Gaussian fields can indeed produce straight structural supports. This phenomenon of oblate spheroidal Gaussian fields combining to form straight structural components can be more intuitively observed in the Gaussian field surface shown in Fig. \ref{fig9}(e).  In particular, when the ends of neighboring Gaussians overlap with each other, their superposition lifts the field to a near-uniform and ridge-like height map; after thresholding, this ridge is extracted as a straight, truss-like member. This confirms the GET method's superior flexibility in representing both straight and curved boundaries, without requiring additional adjustments or Boolean operations.

\subsubsection{2D Compliant mechanism example}
\begin{figure}[h]
	\centering
		\includegraphics[scale=.225]{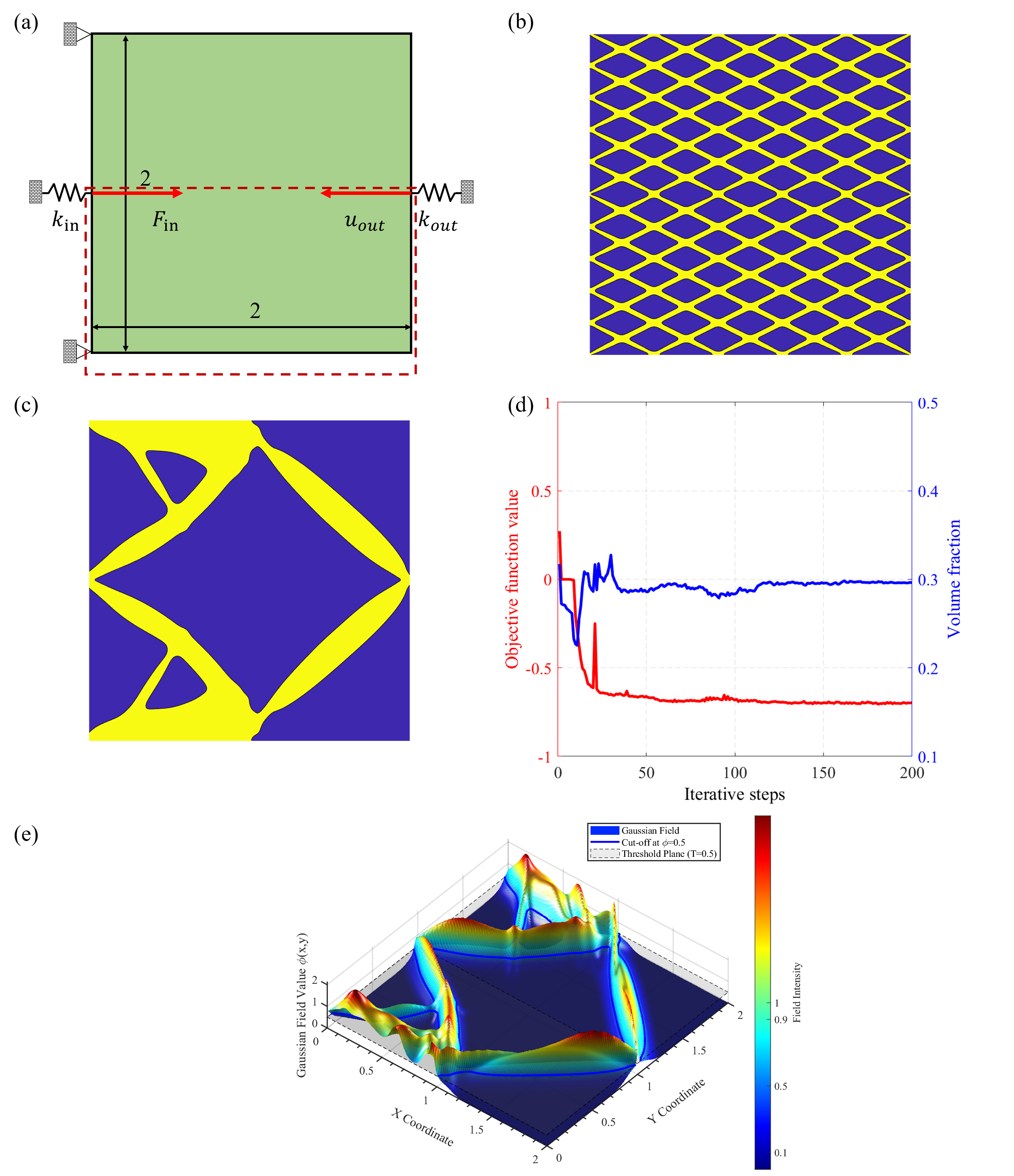}
	\caption{(a) Boundary conditions of the 2D compliant mechanism example; (b) Initial layout of Gaussian fields; (c) Optimized configuration; (d) Iterative convergence curve; (e) Surface plot (height map) of Gaussian fields in the optimized configuration}
	\label{fig10}
\end{figure}
The final benchmark is the classical compliant mechanism design problem, which is widely used in microelectronics. The design domain and boundary conditions are illustrated in Fig.~\ref{fig10}(a). A unit actuation force $F_{\text{in}}=1$ and an input spring ($k_{\text{in}}=0.1$) are applied at the midpoint of the left edge (input port). At the midpoint of the right edge (output port), another spring of stiffness $k_{\text{out}}=0.1$ is attached. Due to symmetry, only half of the $2\times 2$ square domain is optimized, discretized by a $200\times 100$ FE mesh. The initial layout employs $5\times5\times2=50$ Gaussian fields, i.e. $50\times5 = 250$ design variables; see Fig.~\ref{fig10}(b).

The goal of topology optimization for compliant mechanisms is to obtain a design that effectively converts input work into output displacement in a prescribed direction. This problem is commonly formulated as the maximization of the Mutual Potential Energy (MPE) \citep{alonso2014topology}. To evaluate the MPE, two load cases are solved: (i) the input force case, defined by 
$\boldsymbol{K}\boldsymbol{u}_1 = \boldsymbol{f}_1$, and (ii) the pseudo-load case, in which a unit force is applied at the output port, 
$\boldsymbol{K}\boldsymbol{u}_2 = \boldsymbol{f}_2$. 
The MPE is then expressed as
\[
J = \boldsymbol{u}_2^{\top}\boldsymbol{K}\,\boldsymbol{u}_1.
\]

After derivation and simplification, the  sensitivity of MPE with respect to design variable $d$ is given by
\[
\frac{\partial J}{\partial d} = -\,\boldsymbol{u}_1^{\top}\frac{\partial \boldsymbol{K}}{\partial \rho_e}\,\boldsymbol{u}_2 \,\frac{\partial \rho_e}{\partial d}.
\]

The optimized design is shown in Fig.~\ref{fig10}(c). A slender hinge-like member emerges naturally, which is a characteristic feature of compliant mechanisms that enables rotational motion with minimal resistance to achieve desired kinematic functionality. Compared with conventional minimum-compliance problems, this compliant mechanism formulation is more challenging due to its non-self-adjoint nature, which increases both modeling and numerical complexity. The convergence history is plotted in Fig.~\ref{fig10}(d), where the objective shows oscillations in the early iterations but eventually stabilizes to a converged design. This case demonstrates that the proposed GET framework can capture delicate details essential for compliant mechanism performance, thereby confirming both the robustness and rationality of the optimization approach.

\begin{figure}[h]
	\centering
		\includegraphics[scale=.24]{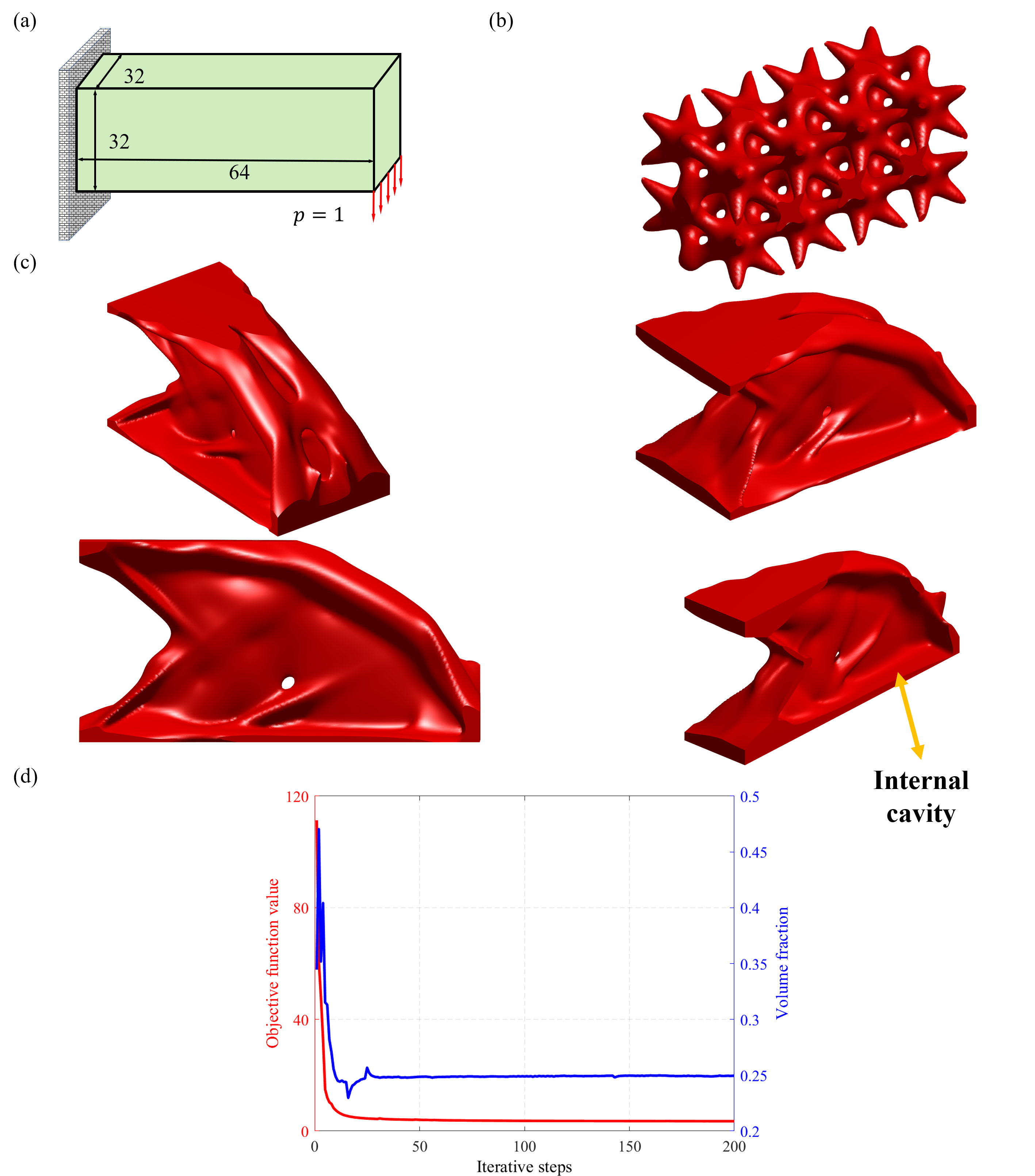}
	\caption{3D cantilever beam obtained by GET method: (a) design domain and boundary conditions; (b) initial design (64 Gaussian functions); (c) optimized design with variable thickness shell forming cavities-shape structure (yellow arrow indicates the cut-open interior); (d) convergence histories of the objective function value and volume fraction.}
	\label{fig11}
\end{figure}

\subsection{3D Numerical Examples}

In this section, three representative 3D benchmark problems are presented to further assess the proposed GET method. 
The 3D cantilever beam and the 3D MBB beam are considered first and then a more complex L-shaped chair problem is studied, which includes both non-designable void and solid domains. 
Together, these examples demonstrate that the proposed method can generate smooth and rational topologies across a range of 3D structural configurations.

\subsubsection{3D cantilever beam example}

A 3D cantilever beam benchmark is employed to validate the proposed methodology in 3D settings. The design domain, loading, and boundary conditions are illustrated in Fig.~\ref{fig11}(a). The cuboidal domain measures $64 \times 32 \times 32$ and is discretized with an $80 \times 40 \times 40$ FE mesh. A uniformly distributed vertical line load of intensity ($p_z = 1$ N) is applied along the bottom-right edge of the free end face, while the left-end face is fully fixed. The volume fraction constraint is prescribed as $0.25$. 

Figure~\ref{fig11}(b) shows the initial design, consisting of $4 \times 2 \times 2 \times 4 = 64$ Gaussian functions, corresponding to 576 design variables (64 Gaussians $\times$ 9 parameters each). Symmetry reduction is not applied here; instead, the full domain is optimized. With symmetric boundary conditions and an initially symmetric layout, the optimization yields a fully symmetric configuration, further confirming the correctness and stability of the 3D implementation.

The optimized topology is presented in Fig.~\ref{fig11}(c). In terms of structural characteristics, the final configuration develops into a classical variable-thickness shell enclosing an internal cavity. By cutting open the structure, it is observed that both the interior and exterior surfaces of the shell are reinforced by protrusions formed through the superposition of Gaussian fields. These protrusions act like smooth stiffeners, enhancing structural rigidity while maintaining continuous transitions. The convergence histories in Fig.~\ref{fig11}(d) show that the compliance $c$ (red line) decreases steadily while the volume fraction $v_f$ (blue line) remains within the prescribed bound.

\subsubsection{3D MBB beam example}
\begin{figure}[h]
	\centering
		\includegraphics[scale=.24]{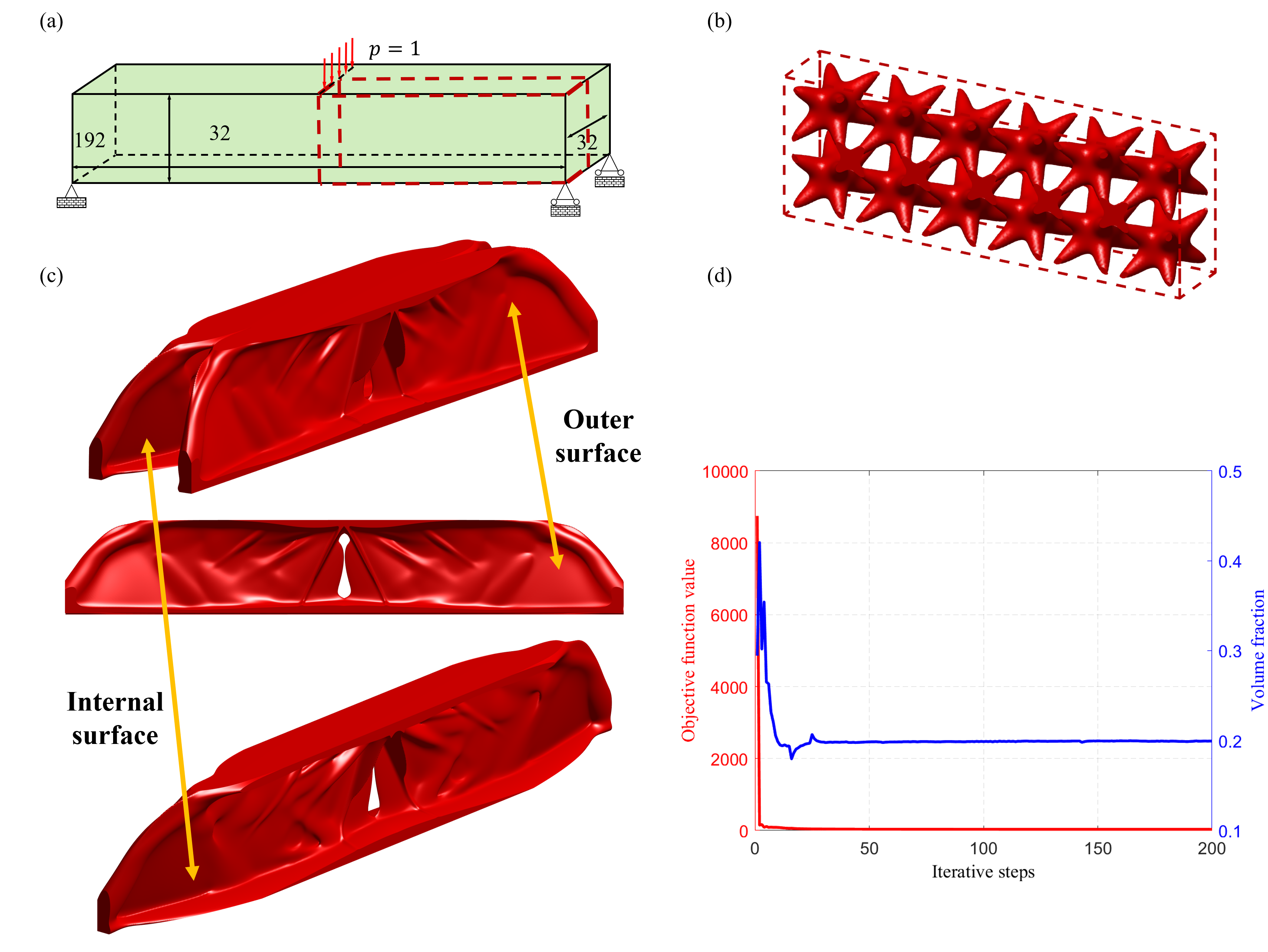}
	\caption{3D MBB beam obtained by GET method: (a) design domain and boundary conditions; (b) initial design of the quarter domain (48 Gaussian functions); (c) optimized design (yellow arrow indicates the cut-open interior); (d) convergence histories of the objective function value and volume fraction.}
	\label{fig12}
\end{figure}

The classical MBB beam problem is investigated in three dimensions as well. 
As shown in Fig.~\ref{fig12}(a), the cuboidal design domain has dimensions $192 \times 32 \times 32$, with the two bottom corner nodes at the left end fully fixed and those at the right end fixed in $y$ and $z$. A uniformly distributed pressure load ($p = 1$) is applied vertically downward along the centerline of the top line at mid-span. The volume fraction constraint is set to $0.2$. Due to the symmetry of the problem setting, only one quarter of the structure discretized with a $120 \times 20 \times 40$ FE mesh is analyzed, as outlined by red dashed line in Fig.~\ref{fig12}(a).

The initial design for the quarter domain is shown in Fig.~\ref{fig12}(b). 
It consists of $6 \times 1 \times 2 \times 4 = 48$ Gaussian functions, corresponding to $48 \times 9 = 432$ design variables.

The optimized topology is presented in Fig.~\ref{fig12}(c). Because the bending moment is relatively large near the mid-span, the cross-sections evolve naturally into an I-shape configuration, effectively resisting bending. 
From the bottom corner supports toward the mid-span, Gaussian fields merge to form a variable-thickness shell, while local protrusions of Gaussian functions create smooth rib-like stiffeners distributed across both inner and outer surfaces.

Figure~\ref{fig12}(d) shows the stable convergence histories. Despite the relatively sparse parameterization (only one layer of Gaussian groups across the beam width), the design evolves from an initial truss-like layout into a variable-thickness stiffened shell, which represents two fundamentally different structural configurations. Moreover, both the global shell surfaces and the local stiffening ribs exhibit smooth, filleted transitions, underscoring the capability of the GET method in capturing complex 3D structural features.

\subsubsection{3D L-shaped chair example}
\begin{figure}[h]
	\centering
		\includegraphics[scale=.22]{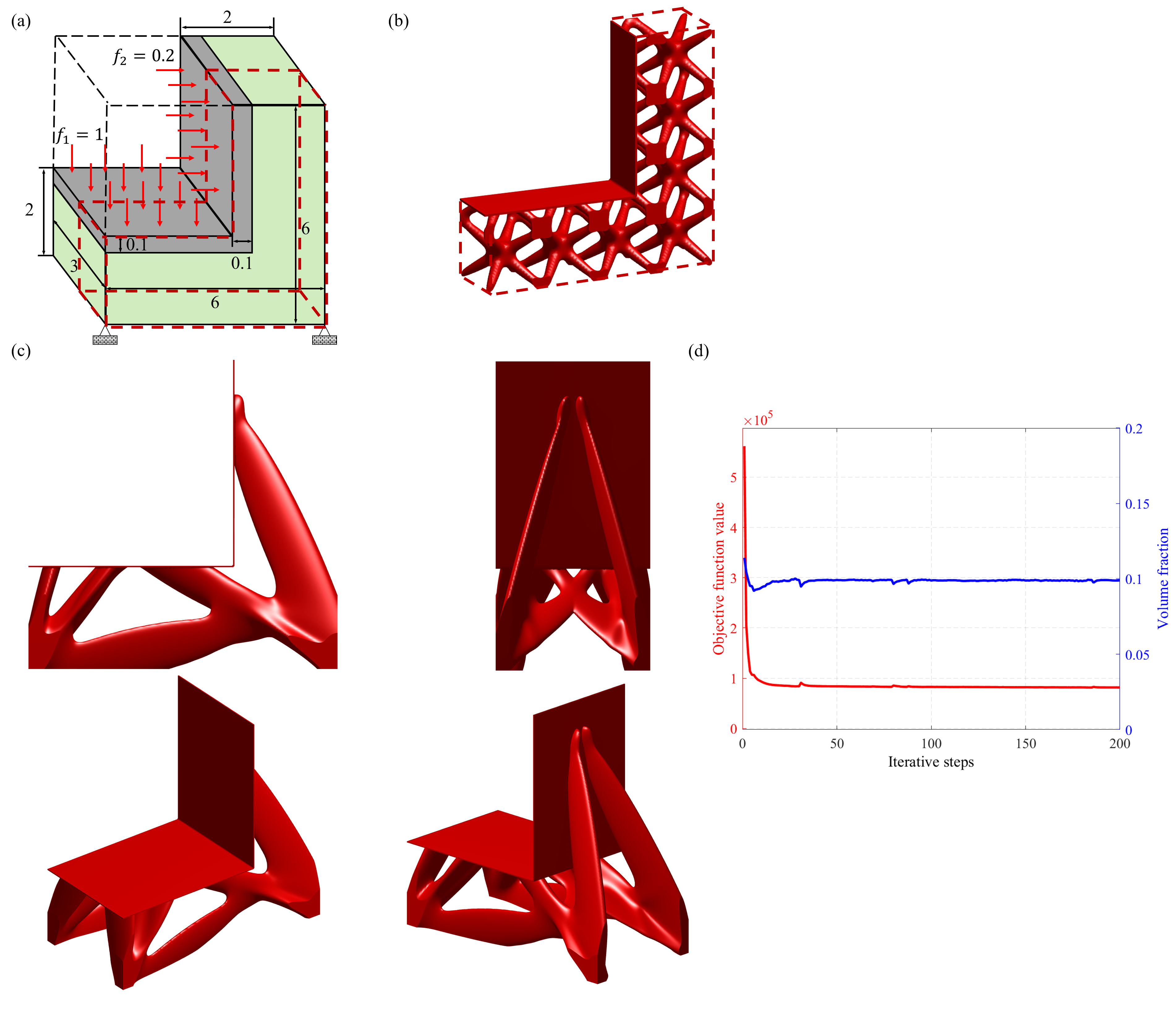}
	\caption{3D L-shaped chair obtained by GET method: (a) design domain and boundary conditions; (b) initial design (56 Gaussian functions); (c) optimized design with chair-like structure; (d) convergence histories of the objective function value and volume fraction.}
	\label{fig13}
\end{figure}

The 3D L-shaped chair problem is considered as a more complex 3D example involving both non-designable void and solid regions. 
As illustrated in Fig.~\ref{fig13}(a), the overall design domain measures $6 \times 3 \times 6$.  Within this domain, a void region of size $4 \times 3 \times 4$ is prescribed at the top-left corner and excluded from optimization. 
In addition, a thin plate subdomain with 0.1  thickness is designated as a frozen region, consistently maintained as solid material throughout the process to ensure reliable load transfer. Surface loads of $f_1 = 1.0$ and $f_2 = 0.2$ are applied to the shaded regions and 
the four bottom corner nodes of the design domain are fully fixed. Under symmetric boundary conditions, only half of the structure is analyzed, corresponding to the subdomain outlined by the red dashed line. The symmetric design domain is discretized by a $100 \times 25 \times 100$ FE mesh, and the allowable volume fraction is set to $0.1$.

The initial distribution of 48 Gaussian functions (corresponding to $48 \times 9 = 432$ design variables) for the half-domain model is shown in Fig.~\ref{fig13}(b). After 200 iterations, the convergence histories in Fig.~\ref{fig13}(d) indicate a steadily improved objective function while the volume fraction consistently satisfies the prescribed constraint. The optimized topology is presented in Fig.~\ref{fig13}(c), which constructed several effective load transmission paths. Compared with the variable-thickness shell structures obtained in the previous beam examples, the present design evolves into a chair-like configuration with clear and smooth supporting members that sustain both the seat surface and the backrest. This case further verifies the capability of the GET method in its effectiveness in geometric representation and applicability to non-designable domains.

\section{Discussion}\label{SDiscussion}

While the previous sections have demonstrated the effectiveness and scalability of the GET method through a series of 2D and 3D benchmark problems, its performance is still governed by several key parameters. 
Among them, four factors are particularly critical:
\begin{itemize}
    \item the mesh size and mesh independence property of explicit method, which affects computational efficiency
    \item the number of Gaussian functions, which determines the representational richness and structural complexity,
    \item the hyperparameter $\epsilon$, which controls the degree of topological discreteness,
    \item the threshold parameter $T$, which regulates boundary smoothness and transition-zone curvature.
\end{itemize}

To systematically assess the influence of these parameters, detailed studies are conducted in this section using the 2D cantilever beam as a representative example.

\subsection{Analysis of Mesh Independence and Computational Efficiency} \label{s41}

The GET method is an explicit topology optimization algorithm, meaning the design is described by explicit geometric parameters rather than implicitly by mesh densities. In theory, this decouples the optimization from the finite element mesh. However, in practice, our study uses a fixed background mesh for all calculations. The element count can still impact the optimized design’s performance and computational cost. It is therefore important to examine how mesh resolution influences the results and to choose an appropriate mesh size that balances simulation accuracy with computational efficiency.

In this case, 2D cantilever beam examples under identical optimization settings but with three different optimization mesh resolutions (coarse $100\times50$, medium $200\times100$, and fine $1000\times500$ elements) are used to investigate mesh independence and efficiency. After obtaining an optimized design on a given mesh, we keep all design variables and parameters and re-evaluate the same structure on the other two mesh resolutions, which are used for post-optimization analysis and comparison. The optimized density distributions for these three cases are shown (highlighted in red boxes) in Table \ref{tb3}. Across the three mesh resolutions (different columns), the objective values are comparable; however, finer discretizations usually admit a slightly more compliant response, so the objective evaluated on the fine mesh is marginally larger. For a fair and intuitive comparison, each optimized design is re-evaluated on the other two meshes (the other two rows). 

\begin{table}[h]
    \centering
    \caption{Mesh-independence study for GET and MMC across three meshes ($100\times50$, $200\times100$, $1000\times500$). 
Columns show that designs are optimized on the indicated mesh (red boxes). The first three rows show the GET method designs re-evaluated on the listed meshes. The penultimate row shows the MMC-optimized designs for the three mesh sizes. The last row and the third row from the bottom show the optimized structures' von Mises stress fields obtained by the GET and MMC methods, respectively, re-evaluated on the  mesh of the $1000 \times 500$ and locally magnified views of the connection regions.}
    \renewcommand{\arraystretch}{1.5} 
\begin{tabular}{
>{\centering\arraybackslash}0{m{1.82cm}}
>{\centering\arraybackslash}0{m{2cm}}
>{\centering\arraybackslash}0{m{1.6cm}}
>{\centering\arraybackslash}0{m{2cm}}
>{\centering\arraybackslash}0{m{1.6cm}}
>{\centering\arraybackslash}0{m{2cm}}
>{\centering\arraybackslash}0{m{1.6cm}}
}

\toprule
\raisebox{1ex}{Examples} & 
\makecell{Optimized on\\ $100\times50$ mesh} & 
\makecell{Obj.\\(Vol.)} & 
\makecell{Optimized on\\$200\times100$ mesh} & 
\makecell{Obj.\\(Vol.)} & 
\makecell{Optimized on\\$1000\times500$ mesh} &
\makecell{Obj.\\(Vol.)} \\
\midrule

\makecell{Evaluated on \\$100\times50$ \\mesh} & 
\centering\adjustbox{valign=c}{\hspace{-1ex}\includegraphics[width=2.4cm, height=1.35cm, keepaspectratio]{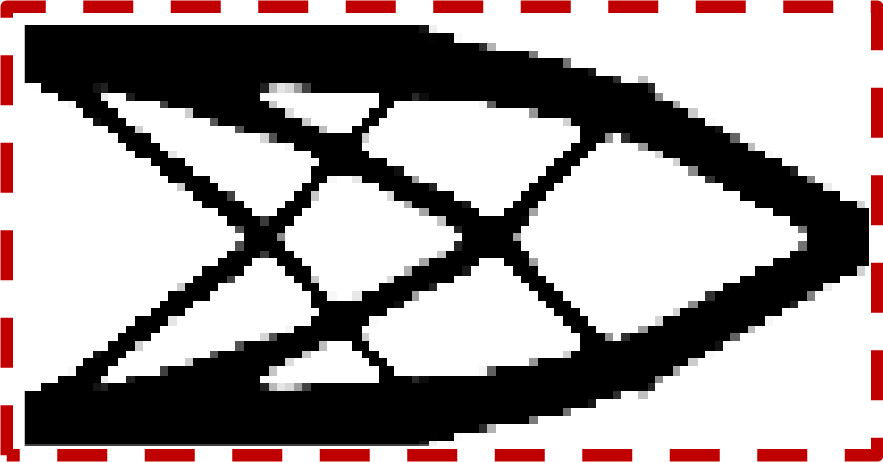}} &   
\makecell{73.4 \\(0.3999)} &
\centering\adjustbox{valign=c}{\hspace{-0.8ex}\includegraphics[width=2.4cm, height=1.35cm, keepaspectratio]{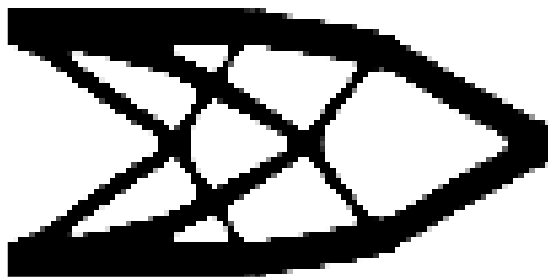}} &   
\makecell{73.1 \\(0.3987)} & 
\centering\adjustbox{valign=c}{\hspace{-1.0ex}\includegraphics[width=2.4cm, height=1.35cm, keepaspectratio]{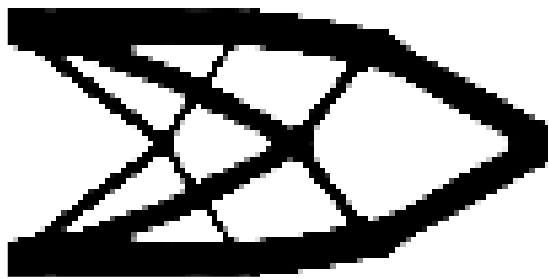}} &   
\makecell{73.2 \\(0.3987)} \\

\makecell{Evaluated on \\$200\times100$ \\mesh} & 
\centering\adjustbox{valign=c}{\hspace{-1ex}\includegraphics[width=2.4cm, height=1.35cm, keepaspectratio]{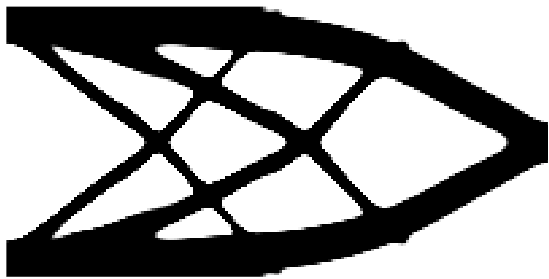}} &   
\makecell{74.1\\(0.4011)} &
\centering\adjustbox{valign=c}{\hspace{-0.8ex}\includegraphics[width=2.4cm, height=1.35cm, keepaspectratio]{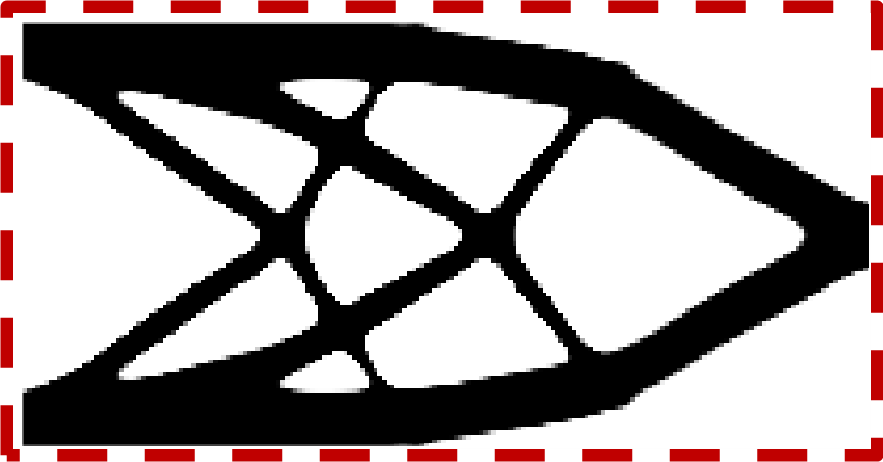}} &   
\makecell{74.0 \\(0.3989)} & 
\centering\adjustbox{valign=c}{\hspace{-1.0ex}\includegraphics[width=2.4cm, height=1.35cm, keepaspectratio]{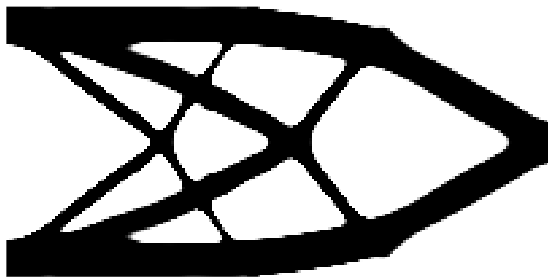}} &   
\makecell{74.0 \\(0.3989)} \\

\makecell{Evaluated on \\$1000\times500$ \\mesh} & 
\centering\adjustbox{valign=c}{\hspace{-1ex}\includegraphics[width=2.4cm, height=1.35cm, keepaspectratio]{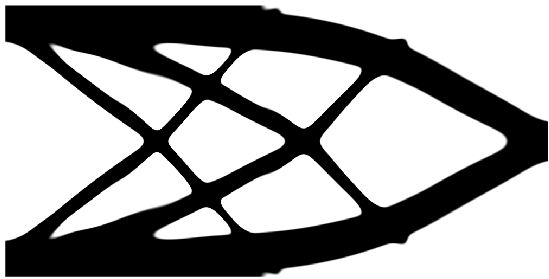}} &   
\makecell{75.2 \\(0.4016)} &
\centering\adjustbox{valign=c}{\hspace{-0.8ex}\includegraphics[width=2.4cm, height=1.35cm, keepaspectratio]{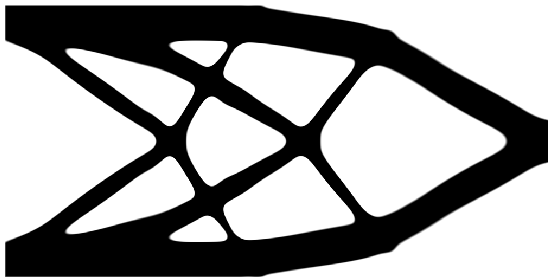}} &   
\makecell{75.0 \\(0.3991)} & 
\centering\adjustbox{valign=c}{\hspace{-1.0ex}\includegraphics[width=2.4cm, height=1.35cm, keepaspectratio]{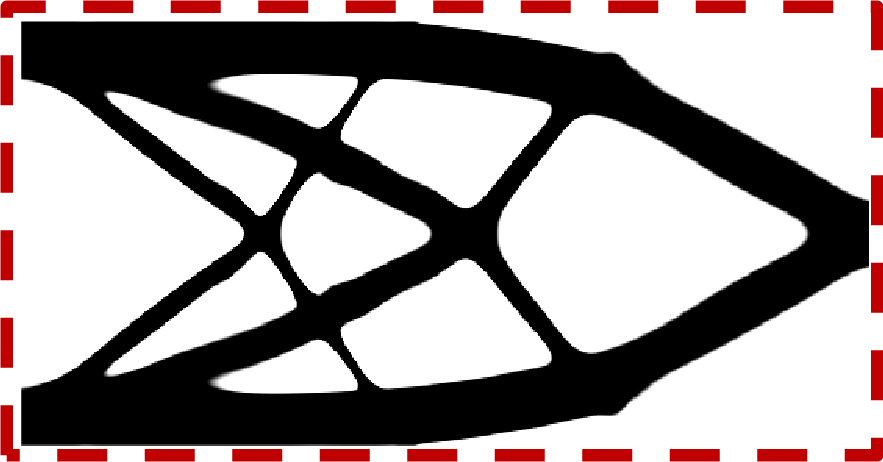}} &   
\makecell{75.0 \\(0.3989)} \\

\makecell{von Mises \\ stress on \\ $1000\times500$ \\ mesh(GET)} &
\multicolumn{2}{>{\centering\arraybackslash}m{4.4cm}}{\adjustbox{valign=c}{\includegraphics[width=6cm, height=1.95cm, keepaspectratio]{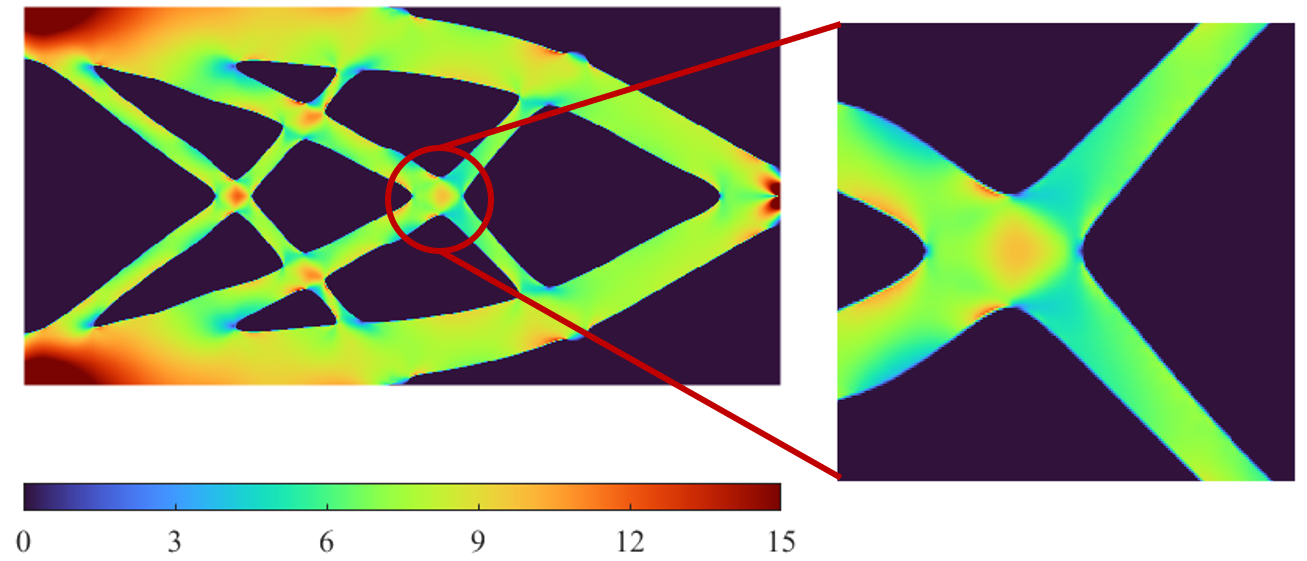}}} &
\multicolumn{2}{>{\centering\arraybackslash}m{4.4cm}}{\adjustbox{valign=c}{\includegraphics[width=6cm, height=1.95cm, keepaspectratio]{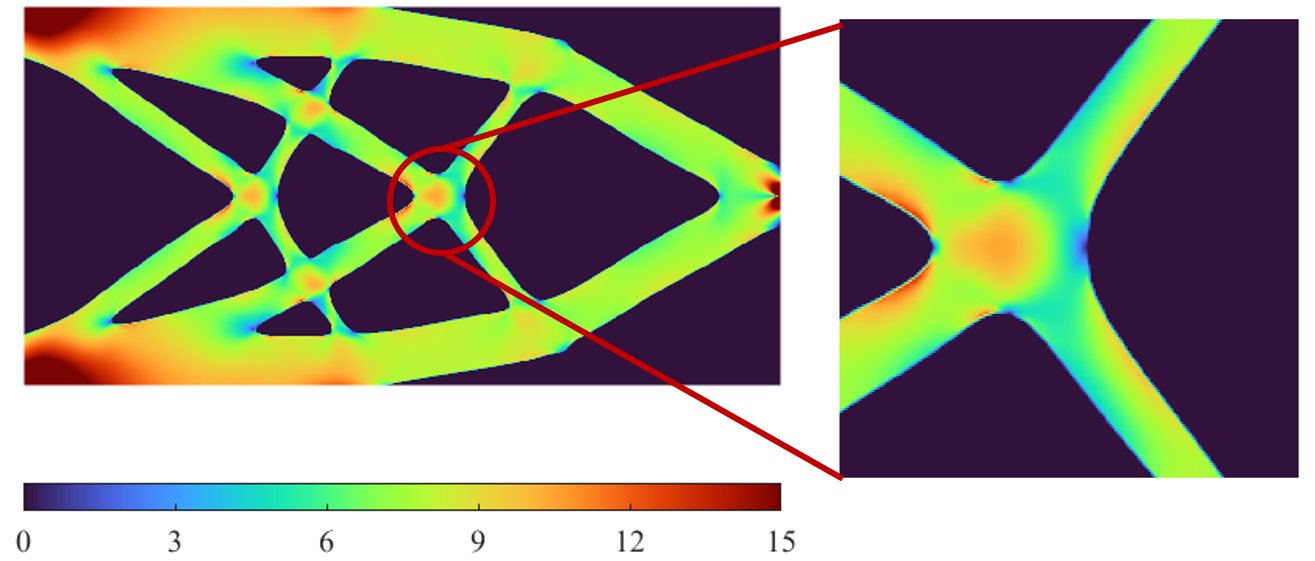}}} &
\multicolumn{2}{>{\centering\arraybackslash}m{4.4cm}}{\adjustbox{valign=c}{\includegraphics[width=6cm, height=1.95cm, keepaspectratio]{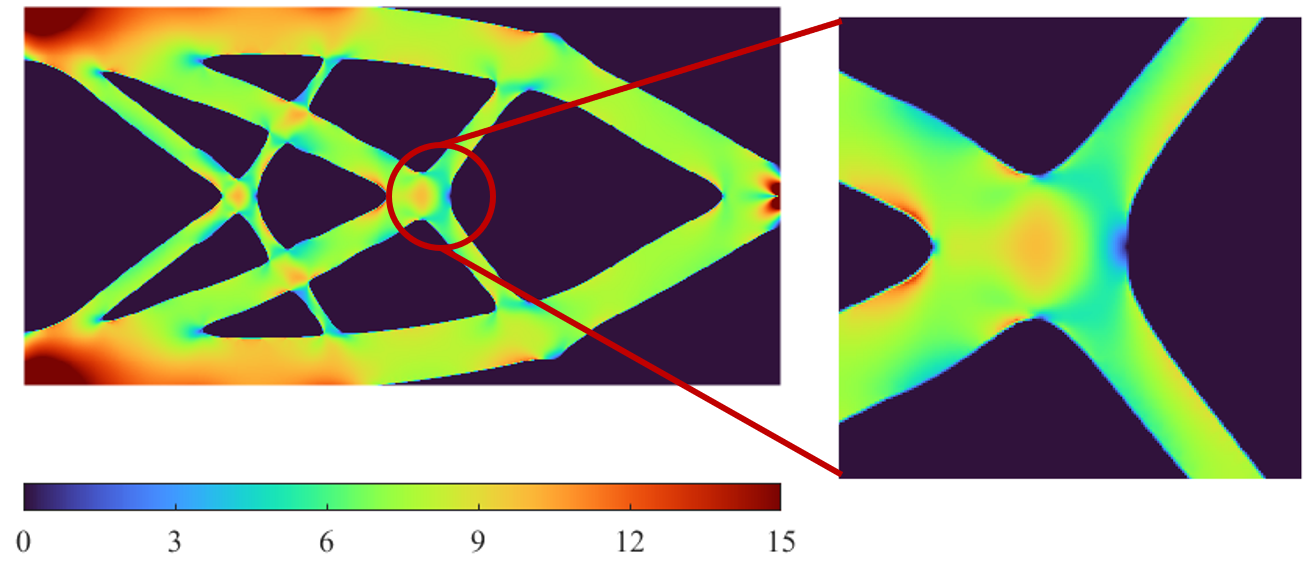}}} \\

\makecell{Design \\ optimized \\ by MMC} & 
\centering\adjustbox{valign=c}{\hspace{-1ex}\includegraphics[width=2.4cm, height=1.35cm, keepaspectratio]{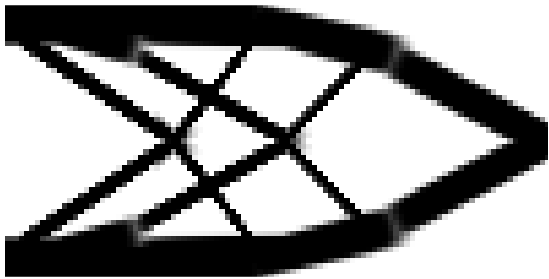}} &   
\makecell{73.2 \\(0.3995)} &
\centering\adjustbox{valign=c}{\hspace{-0.8ex}\includegraphics[width=2.4cm, height=1.35cm, keepaspectratio]{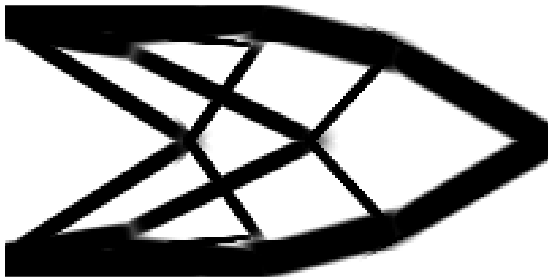}} &   
\makecell{73.4 \\(0.3999)} & 
\centering\adjustbox{valign=c}{\hspace{-1.0ex}\includegraphics[width=2.4cm, height=1.35cm, keepaspectratio]{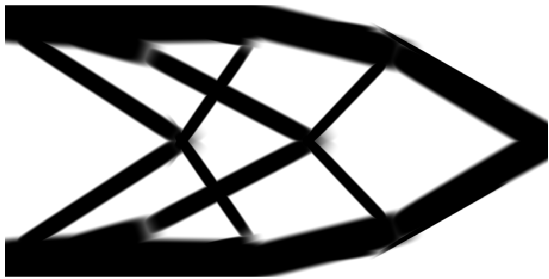}} &   
\makecell{74.8 \\(0.3996)} \\

\makecell{von Mises \\ stress on \\ $1000\times500$ \\ mesh(MMC)} &
\multicolumn{2}{>{\centering\arraybackslash}m{4.4cm}}{\adjustbox{valign=c}{\includegraphics[width=6cm, height=1.95cm, keepaspectratio]{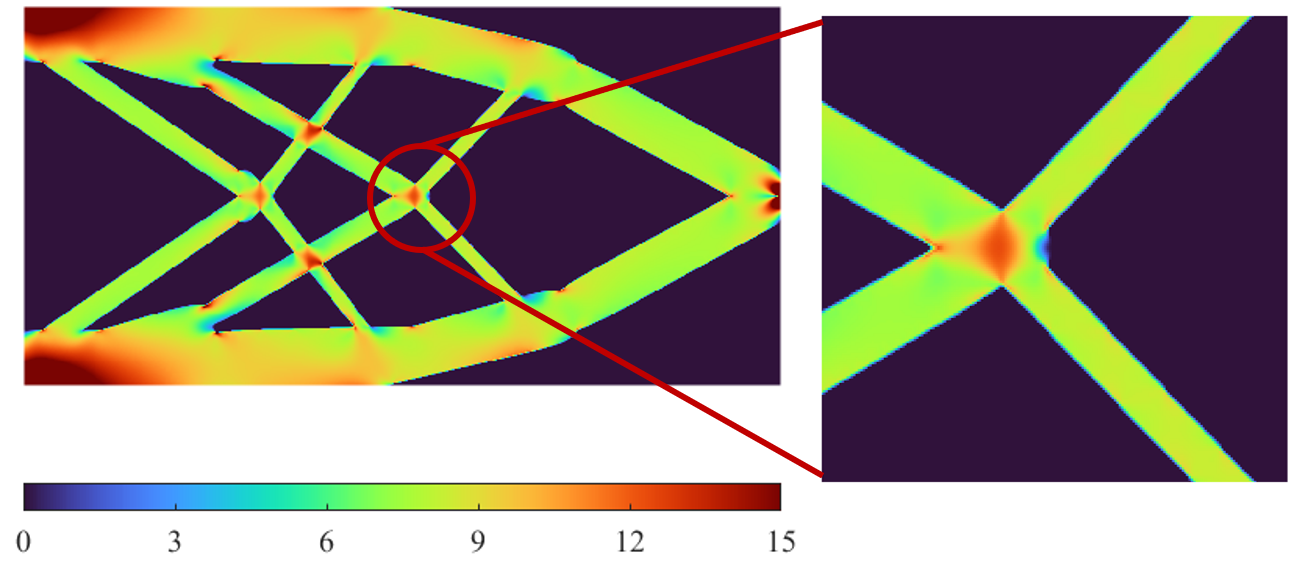}}} &
\multicolumn{2}{>{\centering\arraybackslash}m{4.4cm}}{\adjustbox{valign=c}{\includegraphics[width=6cm, height=1.95cm, keepaspectratio]{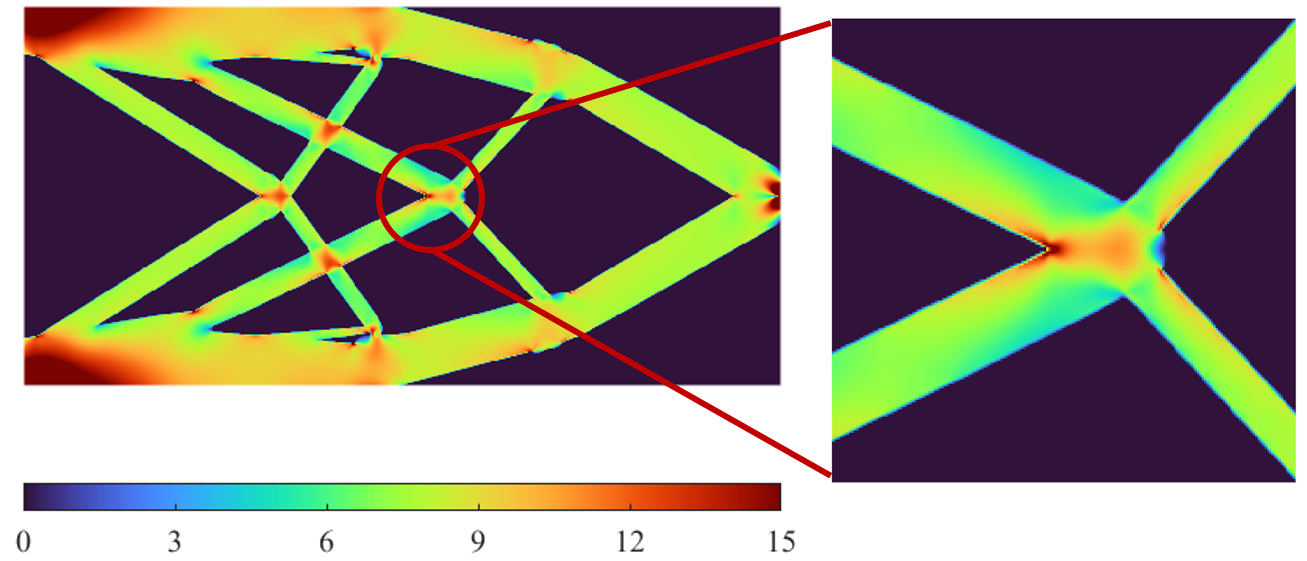}}} &
\multicolumn{2}{>{\centering\arraybackslash}m{4.4cm}}{\adjustbox{valign=c}{\includegraphics[width=6cm, height=1.95cm, keepaspectratio]{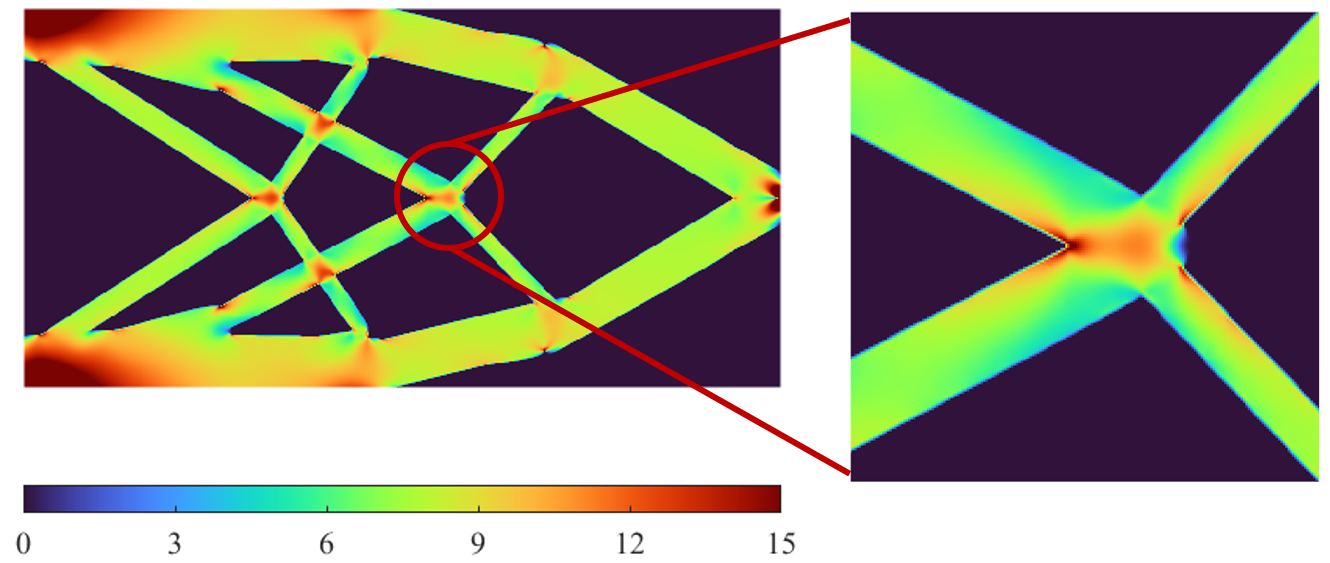}}} \\
        \bottomrule
    \end{tabular}
    \label{tb3}
\end{table}

It can be observed that when evaluated on the same fine mesh, all three designs (optimized from coarse, medium, and fine meshes) achieve very similar objective values and exhibit nearly identical structural configuration, while the design from the coarsest mesh is only marginally suboptimal. The coarsest-mesh design does lack some of the smooth transitional details when viewed on its native coarse mesh. Yet, if we project that coarse design onto a finer mesh, it still presents smooth, filleted structural features. This demonstrates the mesh-independence of the GET representation: the underlying Gaussian design variables define a geometry that remains essentially invariant under mesh refinement or coarsening. This property is particularly useful when a very fine FE mesh is required for accuracy or to capture small boundary details: one can optimize on a coarser mesh for speed, then simply evaluate the final design on a finer mesh to obtain a high-resolution, smooth geometry. All the curved, filleted features introduced by the Gaussian description are preserved with greater clarity on a refined mesh, and the performance difference is negligible. In contrast, attempting a direct optimization on an ultra-fine mesh would be prohibitively expensive in terms of computation. Thus, the GET method inherits the desirable mesh-independence characteristic of explicit topology optimization algorithms, meaning the design parametrization itself is not tied to a particular mesh density.

\begin{figure}[h]
	\centering
		\includegraphics[scale=.6]{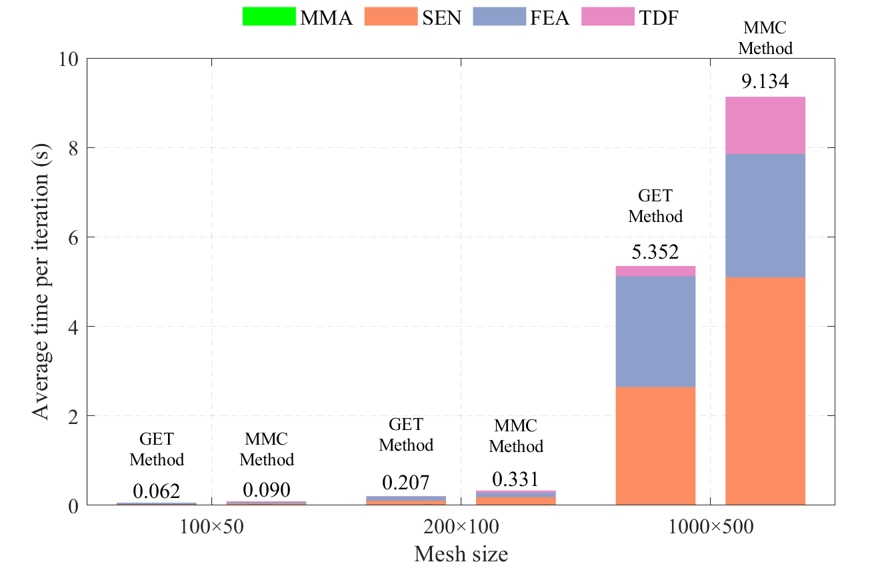}
	\caption{ Average iteration time of GET and MMC methods under different mesh sizes ($100 \times 50$, $200 \times 100$, and $1000 \times 500$). 
The stacked bars indicate the time breakdown of four major computational steps: topology description function (TDF) construction, sensitivity analysis (SEN), finite element analysis (FEA), and MMA updates.}
	\label{fig16}
\end{figure}

This section next examines the computational efficiency of GET in details, breaking down the time spent in each major step of the optimization. The GET method is compared against the MMC approach under the same conditions (using the same number of Gaussian functions or components, a similar initial layout, and the same component activation strategy). The MMC-optimized designs for the three mesh sizes  are included in Table \ref{tb3} (the penultimate row) along with their performance metrics, for reference.
Fig. \ref{fig16} shows the average time per iteration for GET and MMC methods under different mesh resolutions. Each bar is divided into four colored segments representing the time spent in: TDF construction (pink), sensitivity analysis (orange), finite element analysis (blue), and MMA optimizer update (green). GET exhibits consistently lower iteration times than MMC across all mesh sizes. Specifically, both MMC and GET methods spend similar time in the finite element analysis and the MMA update steps, so the efficiency performance gap mainly comes from the topology description function (TDF) construction and sensitivity analysis phases, where GET is faster. There are two primary reasons: the MMC method introduces more design parameters for each component (in this case, MMC had 192 total design variables versus 160 in GET); and the other, more important cause is that the MMC approach requires a Kreisselmeier–Steinhauser (KS) aggregation procedure to merge multiple components, which significantly increases the cost of the TDF and sensitivities. The GET method avoids this overhead by representing the topology as a direct summation of Gaussian fields, eliminating the need for field blending.

It can also be noted that the MMA optimizer update time is negligible in both methods – only 0.0012 seconds per iteration – and it remains roughly constant regardless of mesh size. This is because the number of design variables in these explicit methods is relatively small and fixed, so the cost of solving the MMA does not grow with mesh refinement. In traditional implicit methods, by contrast, the number of design variables equals the number of elements or nodes, which can be tens or hundreds of thousands for fine meshes, causing the optimization update step to scale poorly. The explicit GET formulation avoids this bottleneck. The FEA is the most time-consuming part of each iteration, and its share of the total time increases with mesh density. This aligns with expectation: with mesh refinement, the linear systems become both larger and more poorly conditioned, which makes them harder to solve and increases the computational cost. In contrast, the TDF construction and sensitivity calculation scale approximately linearly with the number of elements, owing to the localized support of each Gaussian and the component activation scheme. Quantitatively, when the mesh is refined from $100\times50$ to $200\times100$ ($4\times$ more elements) and then to $1000\times500$ ($100\times$ more elements), the average iteration time of GET increases by factors of about 3.34 and 85.93, respectively. This scaling is slightly sub-linear with respect to the total element count, indicating good computational efficiency of the implementation.

Finally, we compare the stress distributions of the optimized designs to highlight the benefit of GET’s smooth geometry. The optimized cantilever designs obtained on each mesh (coarse, medium, and fine for both GET and MMC) are projected onto a very fine evaluation mesh of $1000\times500$ for a detailed stress analysis. For clear visual comparison, the von Mises stress fields for all cases are plotted with the same color scale and the maximum value is 15. The critical connection regions (highlighted by circles in the stress plots) reveal a stark difference between GET and MMC designs. The GET-optimized structures, which feature curvature-continuous, filleted connections, show reduced stress concentrations at the joints. The smooth transitions distribute stress more evenly, avoiding high singular peaks. In contrast, the MMC designs have sharp corners and straight junctions, which lead to pronounced stress hotspots  at those connection points. This indicates that beyond achieving comparable compliance performance, the GET method yields structures with improved stress characteristics due to the geometric continuities.

\subsection{Complexity-Gaussian Field Quantity}

\begin{table}[htb]
    \centering  
    \caption{Effect of Gaussian function number on structural complexity, optimized topology, performance and efficiency in the cantilever beam example
}
    \renewcommand{\arraystretch}{3.2}  
    \begin{tabular}{m{2cm} m{3cm} m{3cm} m{2cm} m{4cm}}  
        \toprule
        \multirowcell{1}{Initial Layout\\(Number of\\ Design Variables)} &     
        \multirowcell{1}{Initial\\ Structure} & 
        \multirowcell{1}{Optimized\\ Structure} & 
        \multirowcell{1}{Obj.\\ (Vol.)} & 
        \multirowcell{1}{Computational\\ Cost (s)} \\
        \midrule
        
        \makecell{$1\times1$\\(10)}  &  
        
        \centering\adjustbox{valign=c}{\includegraphics[width=\linewidth, height=1.75cm, keepaspectratio]{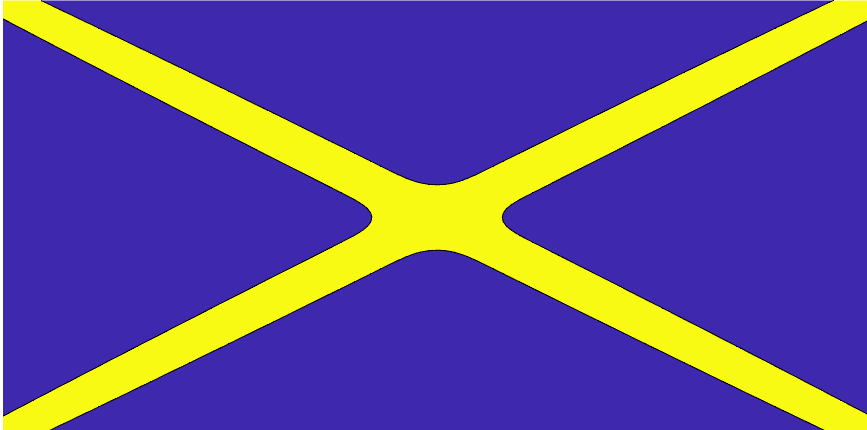}} &         
        \centering\adjustbox{valign=c}{\includegraphics[width=\linewidth, height=1.75cm, keepaspectratio]{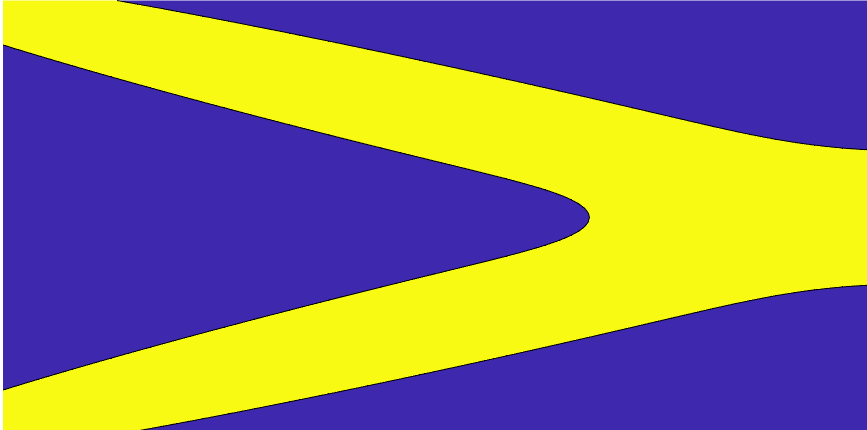}} & 
            \makecell{92.4\\(0.3988)}  & 
            \makecell{Iteration: 0.0996;\\ SEN: 0.0075; FEA: 0.0873;\\ TDF: 0.0009; MMA: 0.0008}  \\

         \makecell{$2\times2$\\(40)}  &
        \centering\adjustbox{valign=c}{\includegraphics[width=\linewidth, height=1.75cm, keepaspectratio]{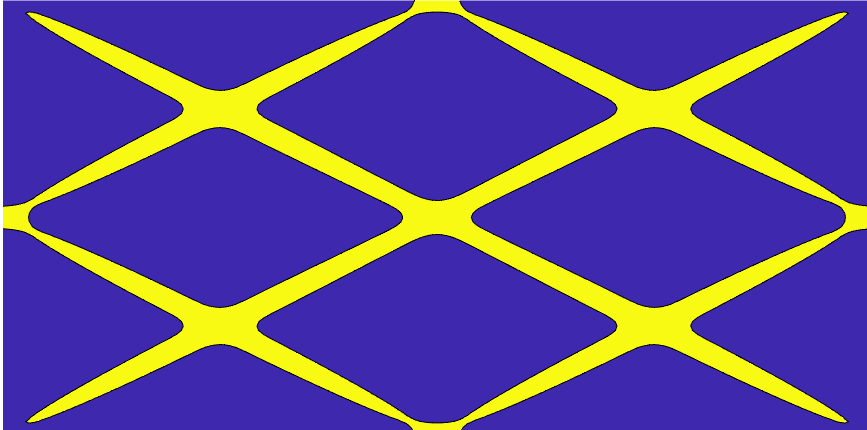}} &         
        \centering\adjustbox{valign=c}{\includegraphics[width=\linewidth, height=1.75cm, keepaspectratio]{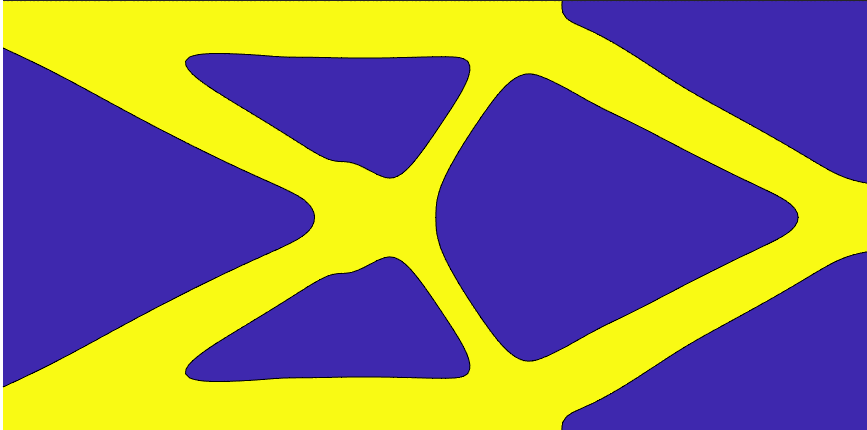}} & 
        \makecell{75.4\\(0.3996)}  & 
        \makecell{Iteration: 0.1204; \\ SEN: 0.0221; FEA: 0.0941; \\ TDF: 0.0032; MMA: 0.0009} \\
        
        \makecell{$4\times4$\\(160)}  &
        
        \centering\adjustbox{valign=c}{\includegraphics[width=\linewidth, height=1.78cm, keepaspectratio]{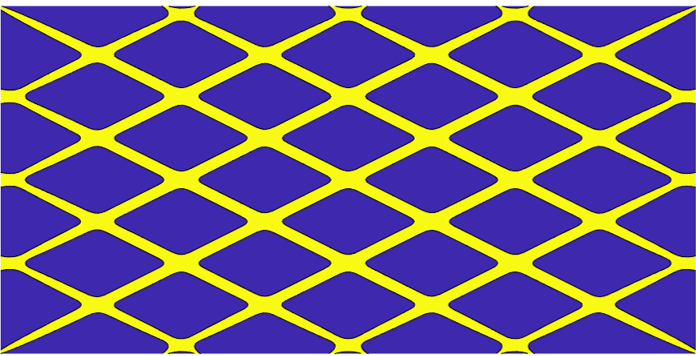}} &         
        \centering\adjustbox{valign=c}{\includegraphics[width=\linewidth, height=1.75cm, keepaspectratio]{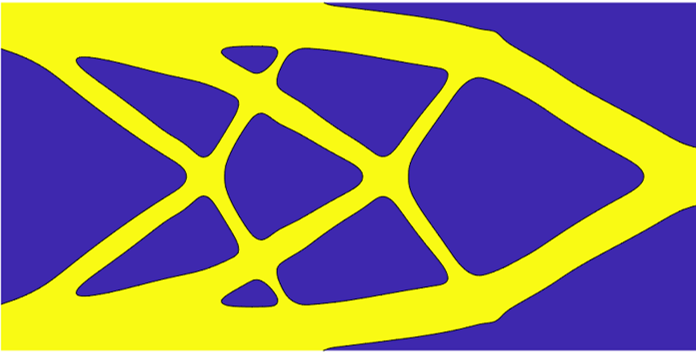}} & 
                \makecell{ 74.0\\(0.3989)}& 
                \makecell{Iteration: 0.2074; \\ SEN: 0.1005 ;FEA: 0.0957; \\ TDF: 0.0100; MMA: 0.0012}  \\ 
        
         \makecell{$6\times6$\\(360)}   &
        
        \centering\adjustbox{valign=c}{\includegraphics[width=\linewidth, height=1.75cm, keepaspectratio]{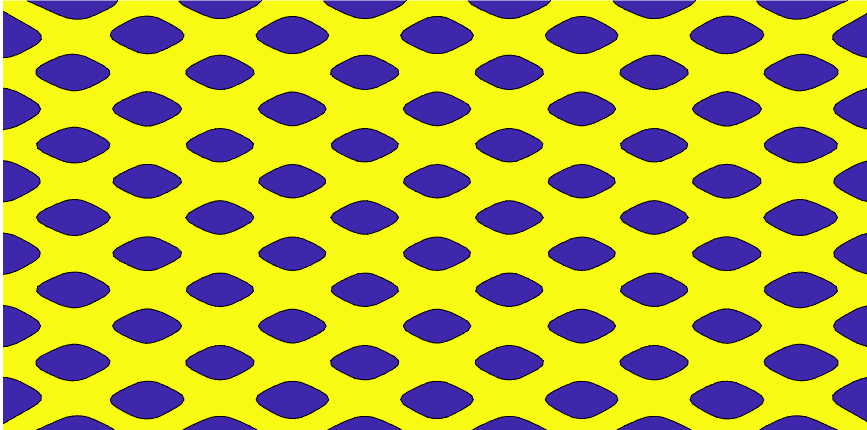}} &         
        \centering\adjustbox{valign=c}{\includegraphics[width=\linewidth, height=1.75cm, keepaspectratio]{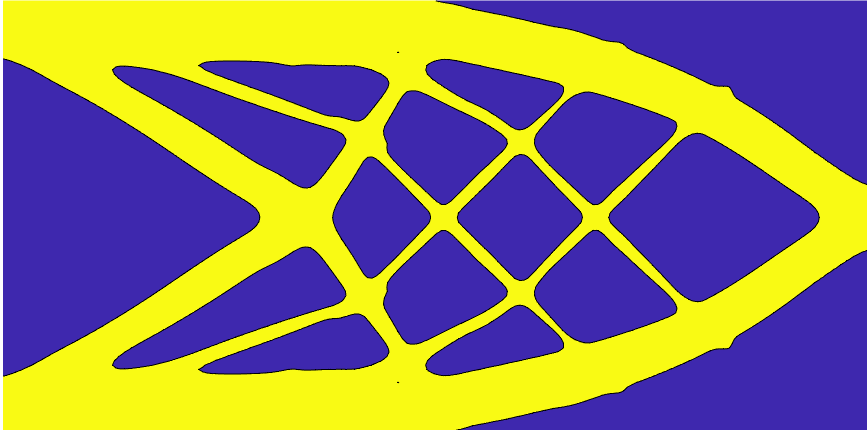}} & 
                        \makecell{72.9\\(0.3999)}& 
                        \makecell{Iteration: 0.3429; \\ SEN: 0.2009; FEA: 0.1234; \\ TDF: 0.0172; MMA: 0.0013}  \\

        \bottomrule
    \end{tabular}
    \label{tb1}
\end{table}

In all numerical examples presented in this paper, the number of Gaussian functions is predefined and remains fixed throughout optimization, meaning that the number of design variables does not change during the process. 
Due to the Gaussian activation scheme (where fields with excessively small covariances are cleared of both material and sensitivity contributions), the number of active Gaussian functions in the final layout is often smaller than the initial allocation. 
This naturally raises two related questions. 
The first is how many Gaussian functions should be distributed, and in what pattern, when tackling a completely new optimization problem. 
This issue is beyond the scope of the present work but may be explored in future studies through adaptive strategies such as Gaussian growth or cloning. 
The second question concerns how the prescribed number of Gaussian functions affects the resulting structural complexity and performance. 
This aspect is systematically investigated here, as illustrated in Table~\ref{tb1}, by varying the number of Gaussian functions in the cantilever beam example.

A straightforward conclusion from these results is that the number of Gaussian functions directly determines the complexity of the optimized design. Fixing the number of Gaussian functions can provide a key function: it offers direct and intuitive control over structural complexity. In engineering design, simpler configurations usually reduce fabrication costs and improve manufacturability, whereas more intricate structures often enhance mechanical performance. Hence, the ability to regulate structural complexity is also one central problem to topology optimization. Implicit methods typically require auxiliary measures such as filter-radius adjustments or genus constraints to achieve this control. In contrast, the GET method inherently regulates complexity, since a small number of Gaussian functions cannot represent highly complex geometries, and the optimization naturally converges to simpler topologies.

The detailed influence of Gaussian field quantity is illustrated in Table~\ref{tb1}, using the cantilever beam problem as an example. 
Here, the initial distribution is refined from $1\times1$ to $6\times6$, with the number of design variables increasing from 10 to 360. 
The third and fourth columns of the table show the initial and optimized structures, respectively. With the $1\times1$ layout, the optimized structure degenerates into a very simple triangular support formed by only two overlapping Gaussians, and the performance deteriorates significantly, with an objective value of 92.4. In 2D problems, structural complexity can be quantified by the genus number (i.e., the number of structural holes)\citep{du2025real}. For this simplest $1\times1$ design case, the genus is zero.

As the number of Gaussian functions increases, the topology evolves into Michell-truss-like patterns with finer structural details and improved performance. In particular, moving from a $1\times1$ to a $2\times2$ layout improves the objective from 92.4 to 75.4, and further refinement continues this trend, reaching 72.9 for the $6\times6$ case. As the number of Gaussian functions increases, the genus number grows correspondingly, rising from 0 to 3, then to 8, and eventually reaching 13 in the denser layouts. This confirms that the number of Gaussian functions provides an effective lever for complexity control, enabling a trade-off between structural complexity and performance.

\begin{figure}[h]
	\centering
		\includegraphics[scale=.2]{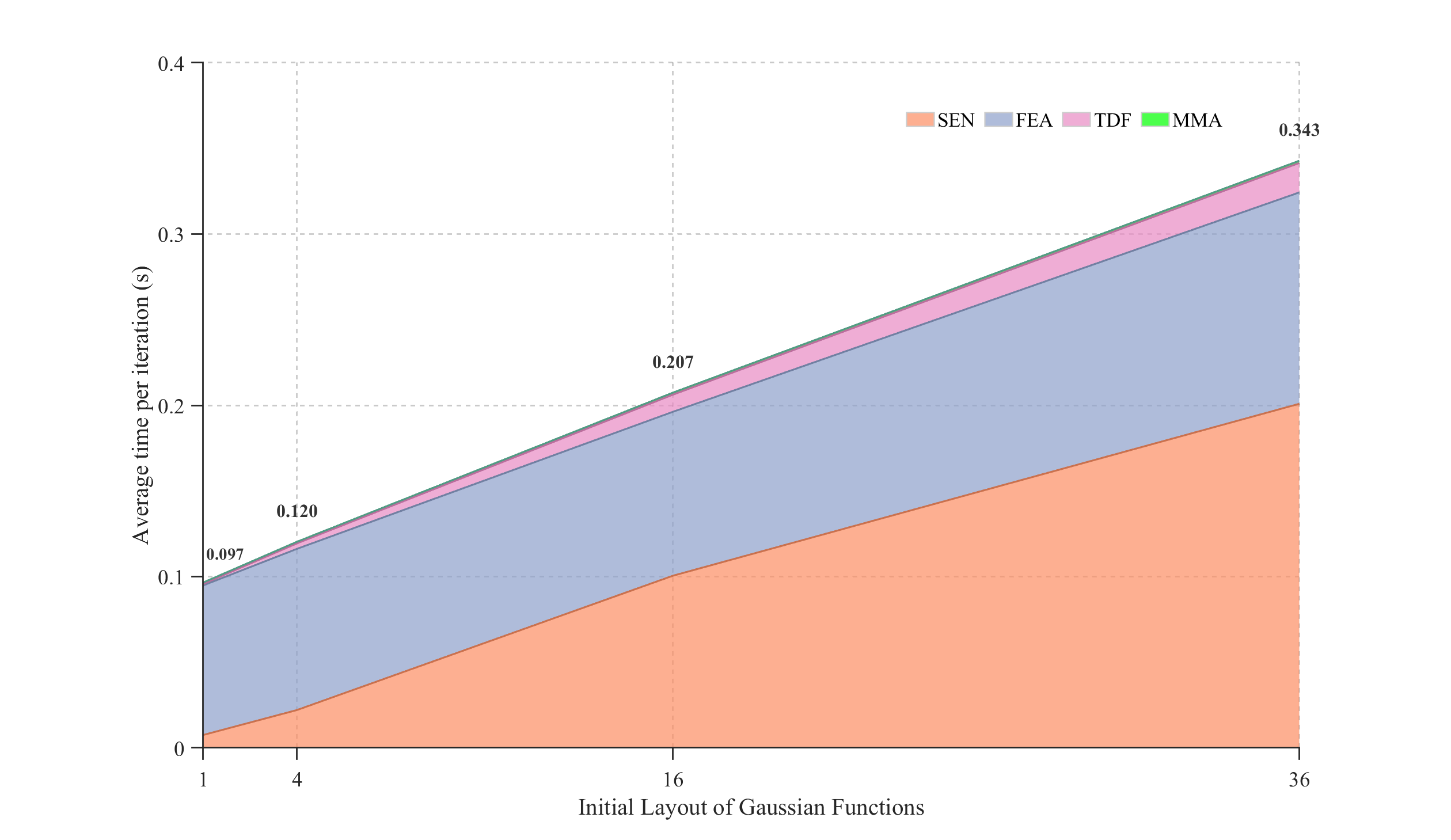}
	\caption{ Average iteration time of GET  methods under different initial layout of Gaussian functions ($1 \times 1$, $2 \times 2$, $4 \times 4$, and $6 \times 6$). 
The stacked area chart indicate the time breakdown of four major computational steps: topology description function (TDF) construction, sensitivity analysis(SEN), finite element analysis (FEA), and MMA updates.}
	\label{fig17}
\end{figure}

 In addition to complexity, the increasing number of Gaussian fields participating in the optimization inevitably sacrifices efficiency. Therefore, a further aspect worth discussing is the influence of the number of Gaussian functions on computational costs. The last column of Table~\ref{tb1} reports the average computation time per iteration for four major components, namely topology description function (TDF) construction, sensitivity analysis (SEN), finite element analysis (FEA), and MMA optimizer updates, with the trends summarized in Fig.~\ref{fig17}. It can be observed that, due to the very small number of design variables involved, the computational cost of the MMA optimizer updates remains negligible even for the larger number of Gaussian functions, consistently below 1\% of the total. Likewise, since the finite element mesh is fixed, the cost of FEA remains almost unchanged. The increase in computation time is mainly attributed to sensitivity analysis and TDF construction. Importantly, however, the growth in cost is sublinear. From the $1\times1$ to the $6\times6$ layout, the number of Gaussian functions increases by a factor of 36, whereas the total computational time increases only by a factor of 3.5; the sensitivity cost rises by 26.8 times and the TDF cost by 19.2 times. This favorable scaling is a direct consequence of the Gaussian activation scheme: in simple layouts, nearly all Gaussians remain active, whereas in more complex topologies redundant Gaussians are automatically deactivated and excluded from SEN and TDF computations. This mechanism accelerates the process and ensures that even highly detailed layouts can still be optimized within an acceptable computational budget.

\subsection{Discreteness-$\epsilon$}
\label{S5.2}

\begin{table}[ht]
    \centering
    \caption{Optimization results for different $\epsilon$ values (0.2, 0.1, 0.02, 0.002): Gray-scale ($\epsilon=\epsilon_0$) and binary ($\epsilon=0$) density maps, with corresponding performance and measure of non-discreteness.}
    \renewcommand{\arraystretch}{2.6}  
    \centering
    \begin{tabular}{m{0.85cm} m{2.5cm} *{2}{m{2.5cm}} *{2}{m{1.6cm}} c}
        \toprule
    \multirowcell{2}{$\varepsilon_0$} & 
    \multirowcell{2}{Optimized\\ Structure} & 
        \multicolumn{2}{c}{\makecell{Density Distribution \\ after Heaviside Function}} & 
        \multicolumn{2}{c}{\makecell{Objective Function Value \\ (Volume Fraction)}} & 
        \multirow{2}{*}{$M_{nd}$} \\
        & &  \centering $\varepsilon=\varepsilon_0$ & \centering $\varepsilon=0.0$ & \centering $\varepsilon=\varepsilon_0$ & \centering $\varepsilon=0.0$ & \\
        \midrule
         \makecell{0.2}  & 
         
        \centering\adjustbox{valign=c}{\includegraphics[width=\linewidth, height=1.15cm, keepaspectratio]{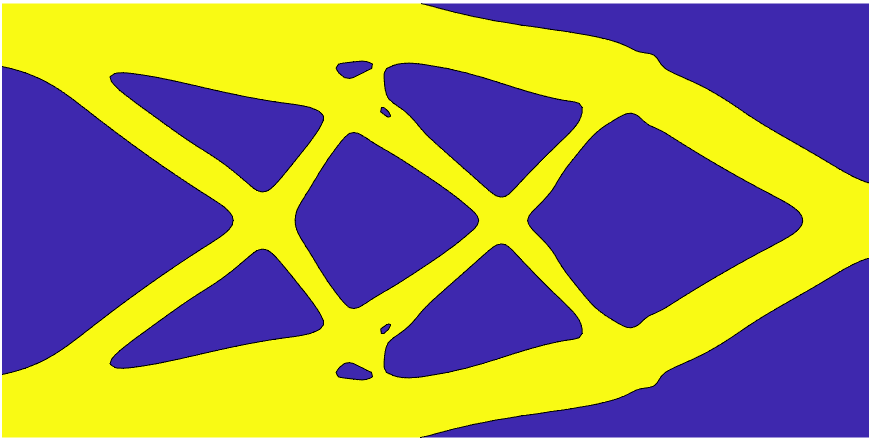}} & 
       \centering\adjustbox{valign=c}{\includegraphics[width=\linewidth, height=1.15cm, keepaspectratio]{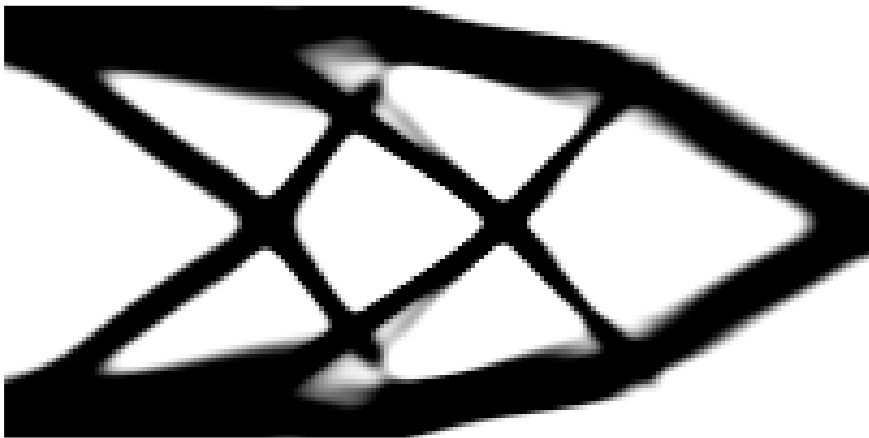}} & 
       \centering\adjustbox{valign=c}{\includegraphics[width=\linewidth, height=1.15cm, keepaspectratio]{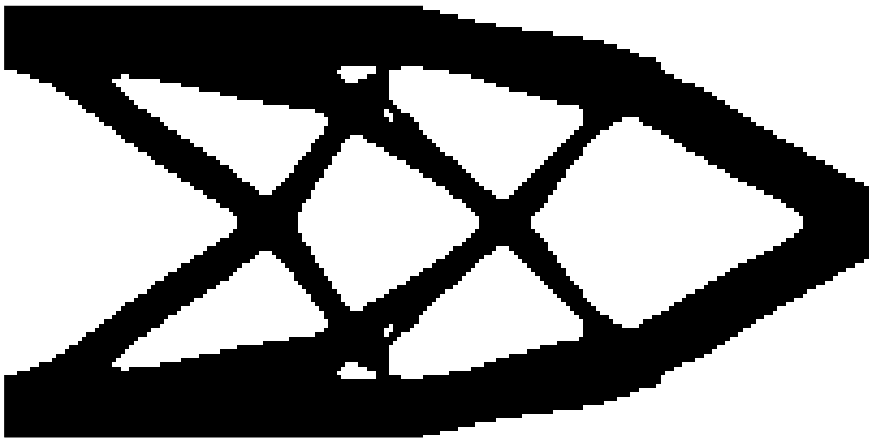}} & 
        \makecell{73.4\\(0.3992)} & 
        \makecell{67.4\\(0.4485)} & 
        10.10\% \\
         \makecell{0.1}  & 
        \centering\adjustbox{valign=c}{\includegraphics[width=\linewidth, height=1.15cm, keepaspectratio]{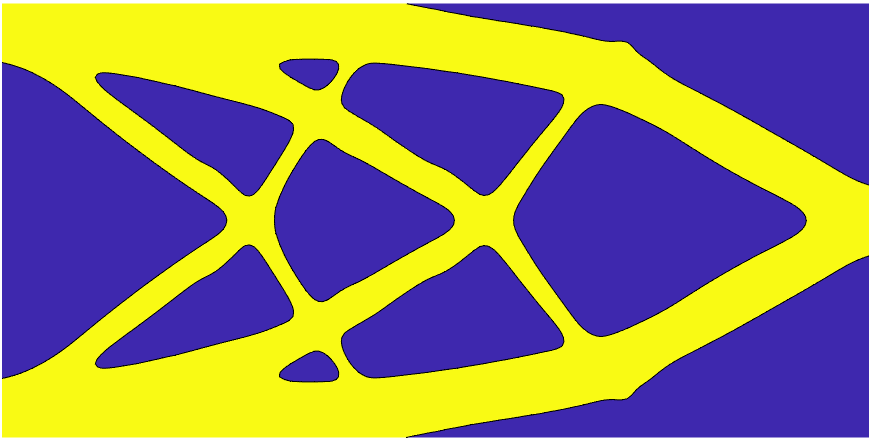}} & 
       \centering\adjustbox{valign=c}{\includegraphics[width=\linewidth, height=1.15cm, keepaspectratio]{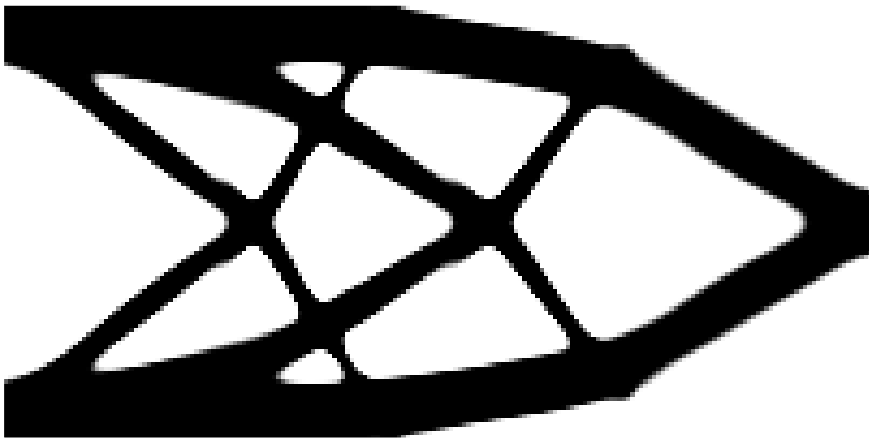}} & 
       \centering\adjustbox{valign=c}{\includegraphics[width=\linewidth, height=1.15cm, keepaspectratio]{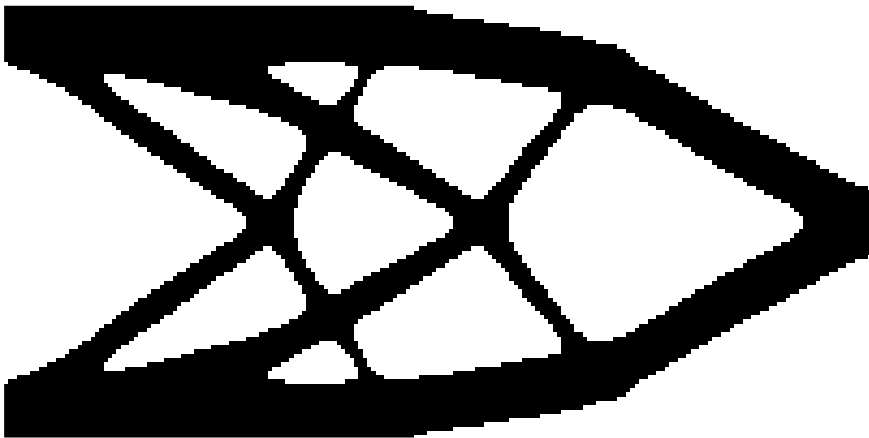}} & 
        \makecell{73.6\\(0.3997)} & 
        \makecell{71.4\\(0.4174)} & 
        3.94\% \\
         \makecell{0.02}  & 
        \centering\adjustbox{valign=c}{\includegraphics[width=\linewidth, height=1.15cm, keepaspectratio]{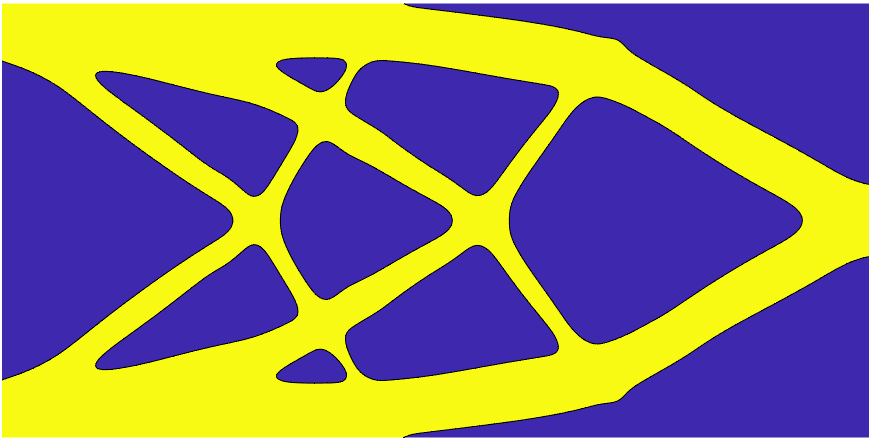}} & 
       \centering\adjustbox{valign=c}{\includegraphics[width=\linewidth, height=1.15cm, keepaspectratio]{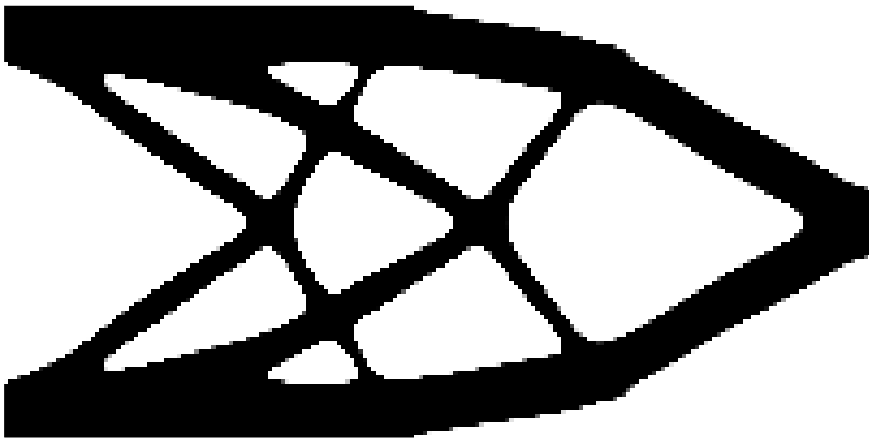}} & 
       \centering\adjustbox{valign=c}{\includegraphics[width=\linewidth, height=1.15cm, keepaspectratio]{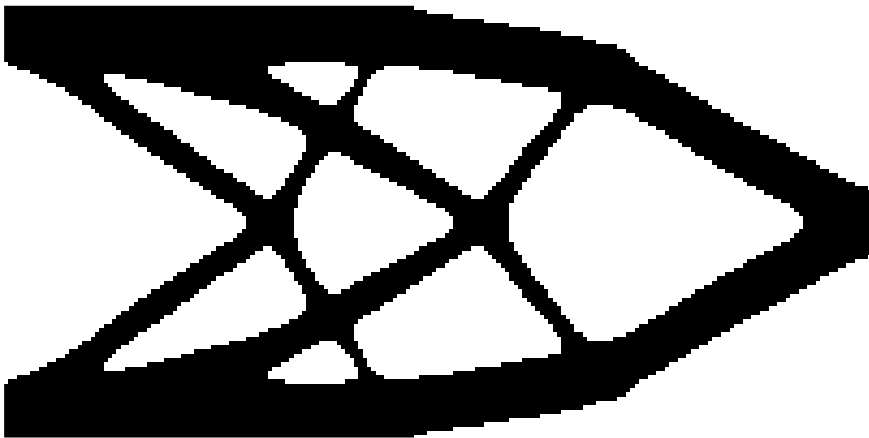}} & 
        \makecell{74.0\\(0.3989)} & 
        \makecell{73.6\\(0.4025)} & 
        0.80\% \\
        \makecell{0.002}  & 
        \centering\adjustbox{valign=c}{\includegraphics[width=\linewidth, height=1.15cm, keepaspectratio]{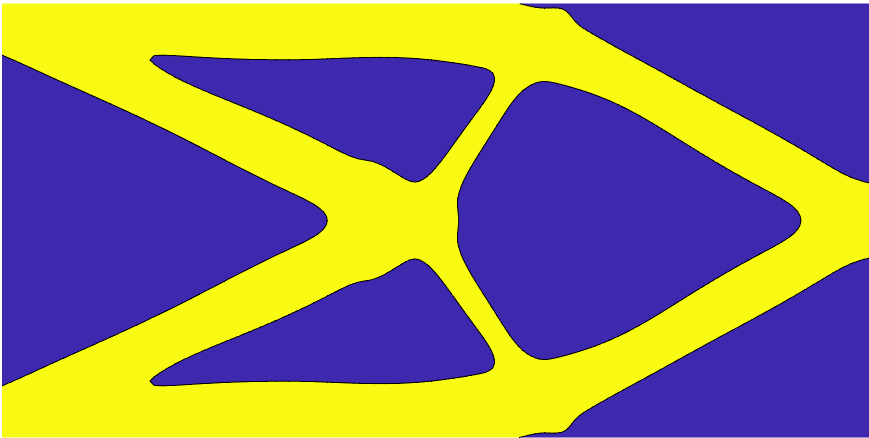}} & 
       \centering\adjustbox{valign=c}{\includegraphics[width=\linewidth, height=1.15cm, keepaspectratio]{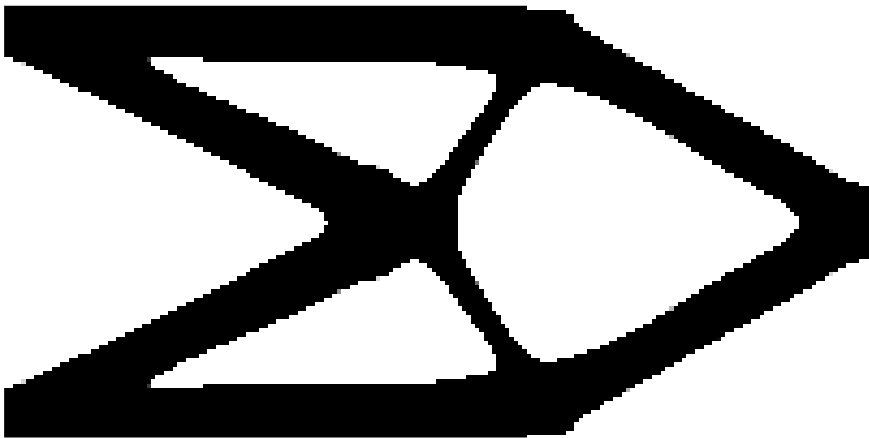}} & 
       \centering\adjustbox{valign=c}{\includegraphics[width=\linewidth, height=1.15cm, keepaspectratio]{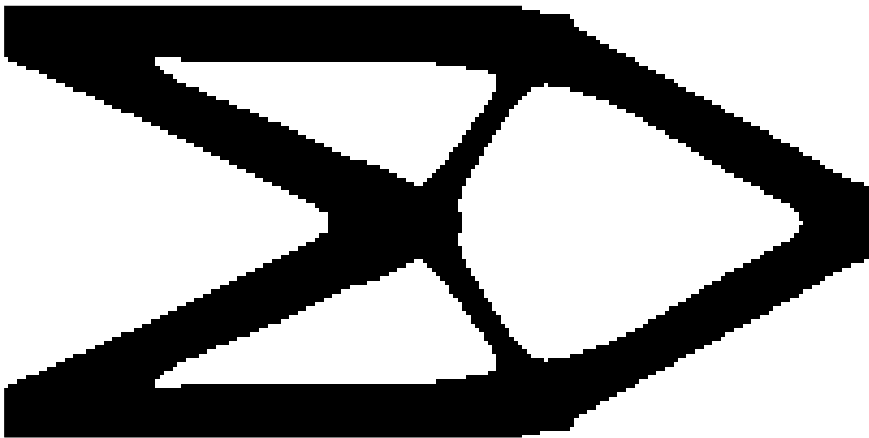}} & 
        \makecell{75.0\\(0.4000)} & 
        \makecell{74.4\\(0.4007)} & 
        0.24\% \\
        \bottomrule
    \end{tabular}
    \label{tb2}
\end{table}

As shown in Fig.~\ref{fig2}, the parameter $\epsilon$ controls the width of the intermediate-density transition band when projecting Gaussian field intensities onto material densities. Larger $\epsilon$ values lead to broader transition zones. Although the GET method is an explicit optimization algorithm, a finite transition region is still required to guarantee gradient existence, which inevitably introduces gray elements. However, an excessive number of gray-scale elements is generally undesirable, since a large proportion of intermediate densities may violate the prescribed volume constraint when extracting the final binary design. To quantify this effect, the so-called measure of non-discreteness ($M_{nd}$)\citep{sigmund2007morphology} is adopted:
\begin{equation}
M_{nd}=\frac{\sum_{e=1}^n 4 \tilde{\rho}_e\left(1-\tilde{\rho}_e\right)}{n} \times 100 \% ,
\label{eq13}
\end{equation}
where $\tilde{\rho}_e$ is the projected density of the $e$-th element. A fully discrete (black-and-white) structure yields $M_{nd}=0\%$, whereas a completely gray structure with uniform $\tilde{\rho}_e=0.5$ yields $M_{nd}=100\%$.

Table~\ref{tb2} summarizes the optimized topologies, density fields, objective values, and $M_{nd}$ for different choices of $\epsilon$ ($0.2, 0.1, 0.02,$ and $0.002$). The second and third column lists the optimized structures and their gray-scale density distributions ($\epsilon=\epsilon_0$). Results show that with $\epsilon=0.2$, the design contains a large proportion of intermediate densities, producing $M_{nd}=10.10\%$. Although the optimization runs with different $\epsilon$ values yield nearly identical objective values, designs containing more intermediate densities with larger $\epsilon$ value appear to perform slightly better during the process. This phenomenon is due to the absence of density penalization in the GET formulation, where gray regions contribute unrealistically to stiffness. In reality, the final designs should be fully discrete. Therefore, for fair comparison, all optimized configurations are further projected with $\epsilon=0$ to eliminate intermediate densities as the final designs and evaluate their objective values and volume fractions.

Notably, for $\epsilon=0.2$, the binary projection results in a volume fraction of 0.4485, exceeding the constraint by 12.3\%, which is unacceptable. However, as $\epsilon$ decreases, $M_{nd}$ is progressively reduced: from 10.10\% ($\epsilon=0.2$), to 3.94\% ($\epsilon=0.1$), to 0.80\% ($\epsilon=0.02$), and finally to 0.24\% ($\epsilon=0.002$). Correspondingly, the binary-projected volume fraction also converges toward the constraint (0.4007 at $\epsilon=0.002$). The case $\epsilon=0.02$, adopted in this study, achieves satisfactory discreteness ($M_{nd}=0.8\%$) while remaining computationally stable. Although smaller values such as $\epsilon=0.002$ further improve discreteness, they risk gradient vanishing in Gaussian fields, which slows convergence and may spuriously deactivate fields with nearly zero sensitivities. Therefore, while the GET method exhibits superior discreteness control compared to SIMP, careful parameter selection is essential. Practical experience suggests that choosing $\epsilon$ within the range $0.01$–$0.1$ balances structural discreteness, numerical stability, and convergence efficiency.

\subsection{Smoothness -$T$}\label{s5}

\begin{figure}[h]
	\centering
		\includegraphics[scale=.3]{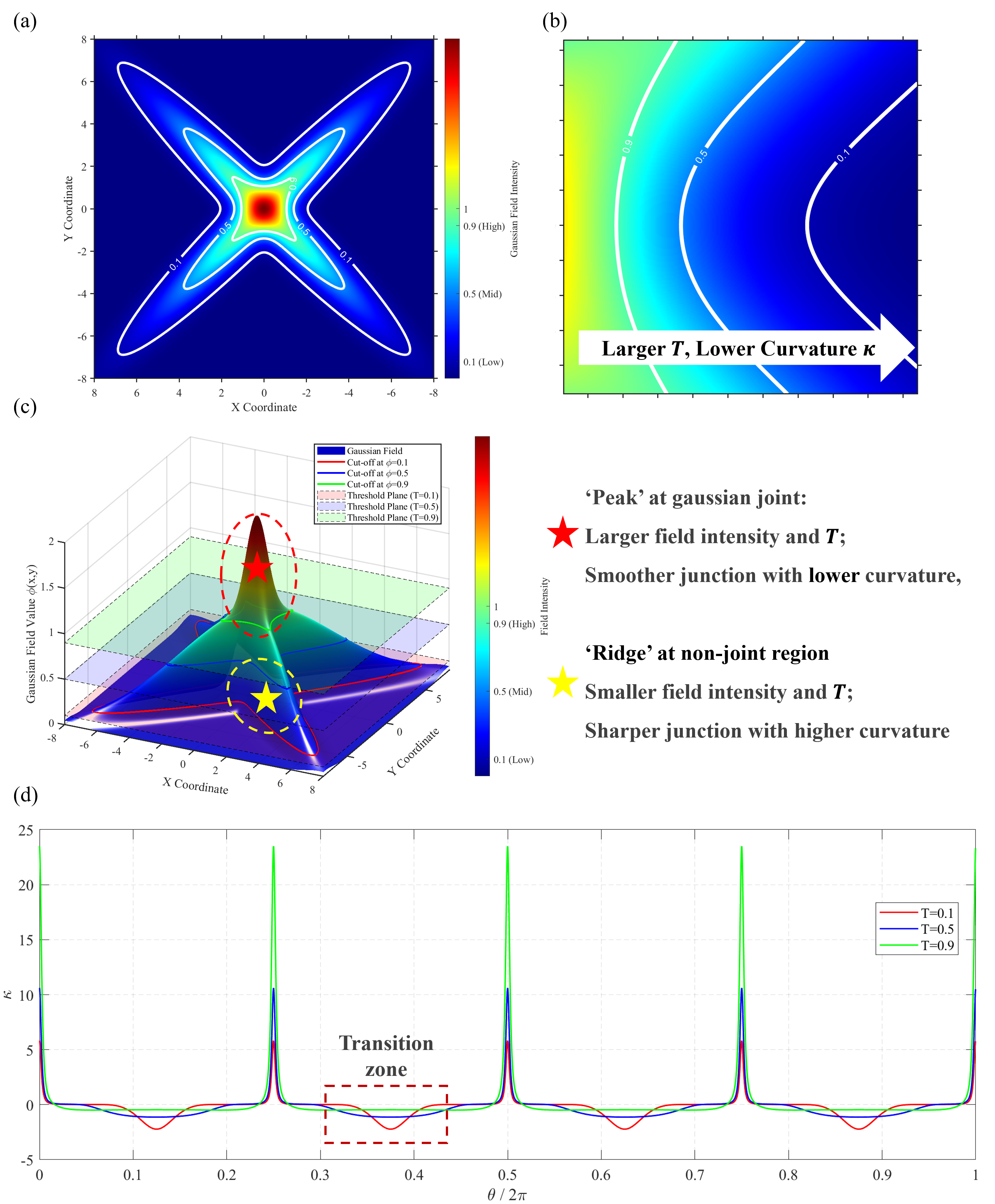}
	\caption{Effect of threshold $T$ on the smoothness of Gaussian field connections, illustrated with two overlapping fields: 
(a) Contour plot showing reduced coverage but smoother boundaries as $T$ increases; 
(b) Magnified connection region where larger $T$ corresponds to smaller curvature $\kappa$; 
(c) Surface plot highlighting a high-intensity peak at the junction (red star, higher curvature, smoother junction) and a ridge-like region (yellow star, lower curvature, sharper junction) truncated under small $T$. (d) Curvature $\kappa$ along the thresholded boundary over one complete traversal, parameterized by the polar angle $\theta$ (abscissa: $\theta/2\pi$).}
	\label{fig14}
\end{figure}

The proposed GET method can generate explicit and smooth topologies, eliminating the need for post-processing. From an engineering standpoint, however, it is often necessary to modulate the degree of smoothness to balance structural performance with manufacturability requirements. The truncation parameter $T$ exactly can provide this control. In Section~4, an intermediate value $T=0.5$ was used, but other values also lead to stable optimization and feasible designs. Importantly, $T$ strongly influences the smoothness and curvature at Gaussian-field junctions, where structural connections form.

As shown in Fig. \ref{fig14}(a) and (c), two Gaussian functions with with rotation angles of $\pi/4$ and $-\pi/4$, respectively, superpose to form the basic unit of the initial layout in the 2D problem. The surface plot (height map) clearly shows that, because the fields are fused by summation, the field intensity at the connections of the Gaussian fields is high; relative to the structural backbone(yellow star region), peak-like features(red star region) appear at this connection region. After truncation by different isocontours ($T = 0.1, 0.5, 0.9$), the resulting contour plot (level-set map) in Fig. \ref{fig14}(a) shows that, as $T$ increases, the extent of the level-set structure decreases noticeably, and the smoothness at the connections improves. Fig. \ref{fig14}(b) further magnifies the structural outline at this location, showing that the curvature of the structure gradually decreases with increasing $T$.

\begin{figure}[ht]
	\centering
		\includegraphics[scale=.275]{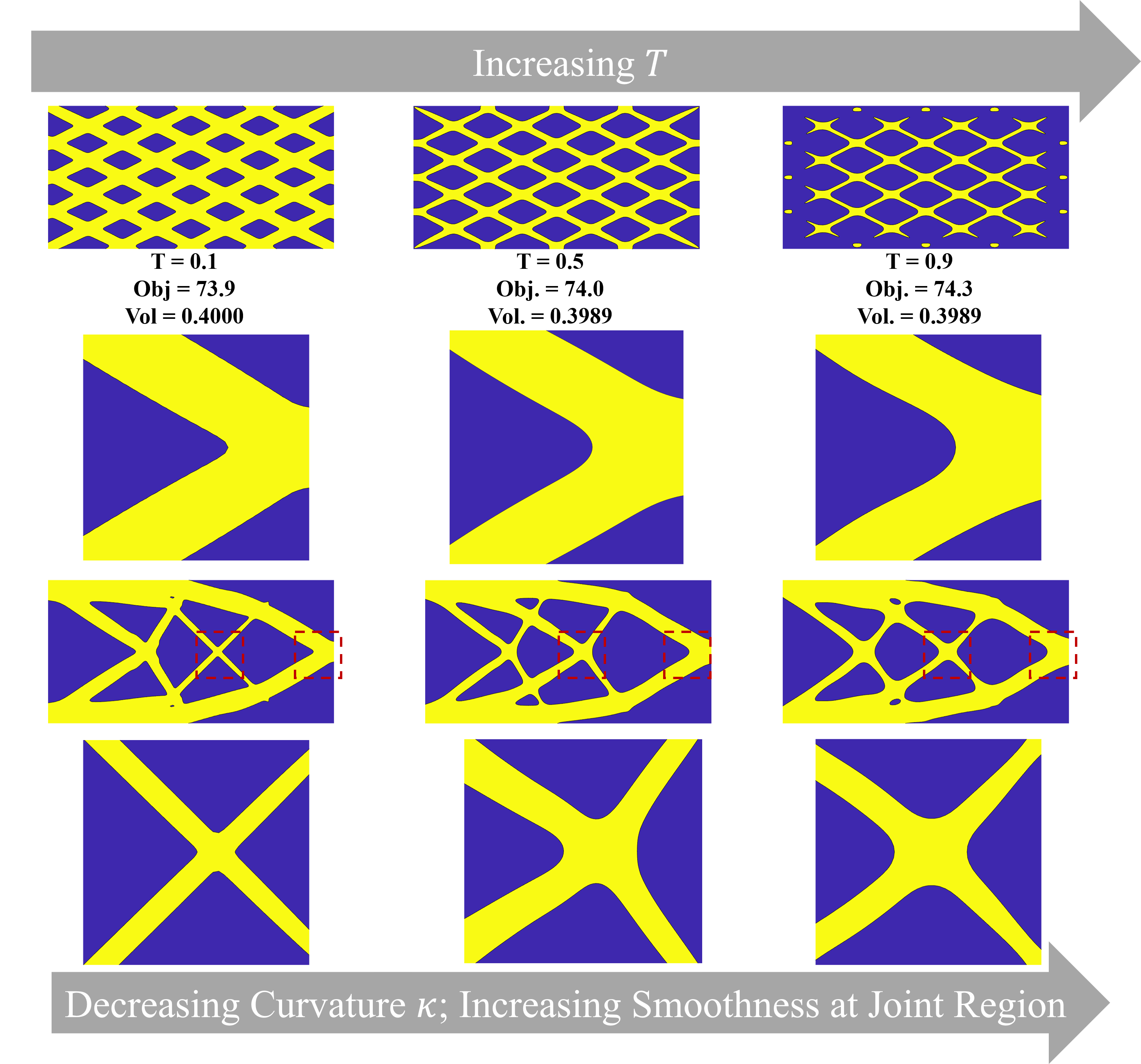}
	\caption{Under the same initial layout, different truncation parameters T (0.1, 0.5, 0.9) yield distinct topological optimization configurations. The red box shows a magnified view of the structural connection area: as T increases, the curvature at the connection decreases, resulting in improved smoothness.}
	\label{fig15}
\end{figure}

The curvature of this level-set boundary ${x: \phi(x)=T}$ is directly determined by the plane curve’s gradient. In particular, suppose that $\boldsymbol{\gamma}(s)$ is a curve in $\mathbb{R}^2$, where $s$ is the arc-length. By defining $\mathbf{n}_s$ as the signed unit normal vector of $\boldsymbol{\gamma}(s)$, the signed curvature is expressed as: 

\begin{equation}
    \frac{d^2 \boldsymbol{\gamma}}{d s^2} = \kappa_s \mathbf{n}_s
\end{equation}

Using the above formulation, the curvature of the closed contours extracted at different cut-off parameters $T$ is computed, and the results are shown in Fig.~\ref{fig14}(d). Negative values correspond to concave transition regions of the structure, while positive values represent convex regions. It can be observed that in the highlighted transition region (red box in Fig.~\ref{fig14}(d)), negative extreme points appear. As $T$ increases, the magnitude of curvature $||\kappa_s||$ decreases (e.g., $T=0.1$: $\kappa_s=-2.23$; $T=0.5$: $\kappa_s=-1.13$; $T=0.9$: $\kappa_s=-0.49$), resulting in smoother geometric transitions.

On the other hand, positive extreme points indicate the curvature at the ends of Gaussian field. In this case, although the curvature increases with larger $T$, making these regions sharper in appearance, such local protrusions are usually suppressed in the formation of the final structure and therefore do not significantly affect the design. Overall, the transition regions of the structure tend to exhibit reduced curvature and hence smoother profiles as $T$ increases. Meanwhile, the structure contours generated by the two Gaussian fields with rotational angles of $\pi/4$ and $-\pi/4$ are not only symmetric with respect to the $x$- and $y$-axes but also invariant under a $90^\circ$ rotation. Consequently, the curvature–angle profiles exhibit perfect symmetry, which further validates the accuracy of the algorithm and implementation.

To further evaluate this effect in the optimization process, Fig.~\ref{fig15} compares optimized designs obtained with different truncation thresholds ($T=0.1,\,0.5,\,0.9$) under the same set of initial design variables. 
It is worth noting that, although the design variables are identical, the thresholding produces distinct initial geometries.  All results satisfy the volume constraint and achieve similar objective values (73.9, 73.6, and 74.3, respectively). While the global topologies performance are comparable, the red insets clearly show local differences at the joints: increasing $T$ reduces curvature and yields smoother transitions. Higher $T$ produces broader radii at joints, alleviating stress concentrations and improving fatigue resistance. 
Nevertheless, excessively large $T$ values may slightly compromise structural performance by suppressing finer load-carrying features. 
Therefore, a balanced choice such as $T=0.5$ is recommended in this study, while adjustments can be made depending on specific design requirements.

\section{Conclusion}\label{SConclusion}

We presented Gaussian Ensemble Topology (GET), an explicit topology-optimization framework that represents geometry by superposed anisotropic Gaussian basis functions and couples this parameterization with a regularized Heaviside projection and analytic sensitivities. Across 2D/3D compliance and compliant-mechanism benchmarks, GET matches MMC objective values while providing curvature-continuous, CAD-friendly boundaries, fewer intermediate densities, and mesh-independent parameterization. The parameters $n, \epsilon, $ and $T$ offer direct control over topological complexity, discreteness, and smoothness. Future work will target stress and fatigue objectives, adaptive Gaussian birth/merge strategies, and GPU-parallel implementations with per-Gaussian analytic sensitivities, enabling larger design spaces and further acceleration.

Looking forward, several promising directions can further extend the applicability of the GET framework. 
First, the GET method can be applied to design problems where smooth geometric transitions are essential, such as stress minimization, fracture-resistance optimization, and complex surface optimization; second, adaptive strategies for cloning or generating Gaussian fields could mitigate the current dependence on preassigned field numbers and enable more flexible design spaces; third, integrating element elimination or adaptive meshing into a Lagrangian formulation would allow the complete removal of weak-density regions, thereby improving computational efficiency and mesh adaptability; 
finally, the independence of Gaussian fields makes the formulation naturally suited for GPU or CPU parallelization, which can be exploited to accelerate both TDF construction and sensitivity analysis in large-scale problems.

Overall, the Gaussian Ensemble Topology method provides a practical and robust approach for explicit topology optimization, bridging structural performance with geometric manufacturability, and laying the groundwork for further extensions to advanced design problems.

\printcredits

\section*{Declaration of Competing Interest}
The authors declare that they have no known competing financial interests or personal relationships that could have appeared to influence the work reported in this paper.

\section*{Acknowledgements}
This research was supported by the the National Natural Science Foundation of China (T2488101), National Research Foundation, Singapore, under the NRF fellowship (Award No. NRF-NRFF17-2025-0006).

\section*{Data availability}
Data will be made available on reasonable request.
\section{Appendix: Exact expressions of $\frac{\partial \phi^i}{\partial d}$} \label{SAppendix}

\subsection{Covariance Matrix Inversion}
The covariance matrix $\boldsymbol{\Sigma}$ decomposes as:
\begin{equation}
\boldsymbol{\Sigma} = \boldsymbol{R}\boldsymbol{S}^2\boldsymbol{R}^\mathrm{T}
\end{equation}
where \(\boldsymbol{R}\) is an orthogonal rotation matrix (\(\boldsymbol{R}^\mathrm{T}\boldsymbol{R} = \boldsymbol{I}\)) and \(\boldsymbol{S} = \text{diag}(\sigma_x, \sigma_y, \sigma_z)\) is a diagonal scaling matrix. Using the inversion property, the inverse covariance matrix is:
\begin{equation}
\boldsymbol{\Sigma}^{-1} = \boldsymbol{R}\boldsymbol{S}^{-2}\boldsymbol{R}^\mathrm{T},
\end{equation}
with \(\boldsymbol{S}^{-2} = \text{diag}(1/\sigma_x^2, 1/\sigma_y^2, 1/\sigma_z^2)\). 

\subsection{Derivatives of the Gaussian design variables}
For the \(i\)-th Gaussian component \(\phi^i(\boldsymbol{x}) = \exp\left(-\frac{1}{2}\boldsymbol{d}^\mathrm{T}\boldsymbol{\Sigma}^{-1}\boldsymbol{d}\right)\) where \(\boldsymbol{d} = \boldsymbol{x} - \boldsymbol{\mu}\), we derive partial derivatives with respect to parameters \(\boldsymbol{d}^i = \{\mu_x^i, \mu_y^i, \mu_z^i, \sigma_x^i, \sigma_y^i, \sigma_z^i, \alpha^i, \beta^i, \gamma^i\}\).

\subsubsection{Gradients with Respect to Mean Parameters }
By the chain rule, the derivative with respect to i-th Gaussian function's $\mu_j$ (Only $\boldsymbol{d}$ depends on $\mu_i$, using \(\boldsymbol{e}_j\) as the \(j\)-th standard basis vector) is:
\begin{equation}
\frac{\partial \phi^i}{\partial \mu_j} = \phi^i \cdot \boldsymbol{e}_j^\mathrm{T}\boldsymbol{\Sigma}^{-1}\boldsymbol{d}.
\end{equation}
Substituting \(\boldsymbol{\Sigma}^{-1} = \boldsymbol{R}\boldsymbol{S}^{-2}\boldsymbol{R}^\mathrm{T}\) gives:
\begin{equation}
\frac{\partial \phi^i}{\partial \mu_j} = \phi^i \cdot \boldsymbol{e}_j^\mathrm{T}\boldsymbol{R}\boldsymbol{S}^{-2}\boldsymbol{R}^\mathrm{T}\boldsymbol{d}
\end{equation}

These are 3D derivations. The 2D case is obtained by suppressing the z-dimension by set $j\in{1,2}$ and drop all terms involving z-dimension.

\subsubsection{Gradients with Respect to Scale Parameters \(\sigma_k\)}

By the chain rule, the derivative with respect to i-th Gaussian function's $\sigma^i$ (Only $\boldsymbol{S}$ in $\boldsymbol{\Sigma}$ is determined  by $\sigma^i$) is: 
\begin{equation}
\frac{\partial \phi^i}{\partial \sigma_j } = -\frac{1}{2}\phi^i \cdot \boldsymbol{d}^\mathrm{T} \frac{\partial \boldsymbol{\Sigma}^{-1}}{\partial \sigma^i}\boldsymbol{d} = -\frac{1}{2}\phi^i \cdot \boldsymbol{d}^\mathrm{T} \boldsymbol{R}\frac{\partial \boldsymbol{S}^{-2}}{\partial \sigma_j}\boldsymbol{R}^\mathrm{T}\boldsymbol{d}
\end{equation}
where,
\begin{equation}
\frac{\partial \boldsymbol{S}^{-2}}{\partial \sigma_k} = \text{diag}\left(0, \ldots, -\frac{2}{\sigma_k^3}, \ldots, 0\right),
\end{equation}
with the non-zero entry only at the \(k\)-th diagonal position. These are 3D derivations. The 2D case is obtained by suppressing the z-dimension by set $k\in{1,2}$ and drop all terms involving z-dimension.

\subsubsection{Gradients with Respect to Euler Angles $\alpha, \beta, \gamma$}

By the chain rule, the derivative with respect to i-th Gaussian function's $\alpha, \beta, \gamma$ (Using $\theta_i, i = 1,2,3$ to represent $\alpha, \beta, \gamma$,only $\boldsymbol{R}$ in $\boldsymbol{\Sigma}$ is determined  by $\theta_i$) is: 

\begin{equation}
\frac{\partial \phi^i}{\partial \theta_j } = -\frac{1}{2}\phi^i \cdot \boldsymbol{d}^\mathrm{T} \frac{\partial \boldsymbol{\Sigma}^{-1}}{\partial \theta^i}\boldsymbol{d} = -\frac{1}{2}\phi^i \cdot \boldsymbol{d}^\mathrm{T} (\frac{\partial \boldsymbol{R}}{\partial \theta_j} \boldsymbol{S}^{-2}\boldsymbol{R}^\mathrm{T}+ \boldsymbol{R}\boldsymbol{S}^{-2}\frac{\partial \boldsymbol{R}^\mathrm{T}}{\partial \theta_j})\boldsymbol{d}
\end{equation}
where,
\begin{subequations}
\begin{align}
\frac{\partial \mathbf{R}_{3D}}{\partial \alpha} &=
\begin{bmatrix}
0 & 0 & 0\\
\cos\alpha\,\sin\beta\,\cos\gamma + \sin\alpha\,\sin\gamma &
\cos\alpha\,\sin\beta\,\sin\gamma - \sin\alpha\,\cos\gamma &
\cos\alpha\,\cos\beta \\
-\sin\alpha\,\sin\beta\,\cos\gamma + \cos\alpha\,\sin\gamma &
-\sin\alpha\,\sin\beta\,\sin\gamma - \cos\alpha\,\cos\gamma &
-\sin\alpha\,\cos\beta
\end{bmatrix},\\[6pt]
\frac{\partial \mathbf{R}_{3D}}{\partial \beta} &=
\begin{bmatrix}
-\sin\beta\,\cos\gamma & -\sin\beta\,\sin\gamma & -\cos\beta\\
\sin\alpha\,\cos\beta\,\cos\gamma & \sin\alpha\,\cos\beta\,\sin\gamma & -\sin\alpha\,\sin\beta\\
\cos\alpha\,\cos\beta\,\cos\gamma & \cos\alpha\,\cos\beta\,\sin\gamma & -\cos\alpha\,\sin\beta
\end{bmatrix},\\[6pt]
\frac{\partial \mathbf{R}_{3D}}{\partial \gamma} &=
\begin{bmatrix}
-\cos\beta\,\sin\gamma & \cos\beta\,\cos\gamma & 0\\
-\sin\alpha\,\sin\beta\,\sin\gamma - \cos\alpha\,\cos\gamma &
\sin\alpha\,\sin\beta\,\cos\gamma - \cos\alpha\,\sin\gamma & 0\\
-\cos\alpha\,\sin\beta\,\sin\gamma + \sin\alpha\,\cos\gamma &
\cos\alpha\,\sin\beta\,\cos\gamma + \sin\alpha\,\sin\gamma & 0
\end{bmatrix}
\end{align}
\end{subequations}

For 2D case, $\boldsymbol{R}_{2D}$ is only determined by $\theta$:

\begin{equation}
\frac{\partial \mathbf{R}_{2D}}{\partial \theta}=\left[\begin{array}{cc}
-\sin \theta & -\cos \theta \\
\cos \theta & -\sin \theta
\end{array}\right]
\end{equation} 

\bibliographystyle{unsrtnat}

\bibliography{ref}

\end{document}